\documentclass{aa}  
\usepackage{marvosym}
\usepackage{graphicx}
\usepackage{txfonts}
\usepackage{lipsum}
\usepackage{upgreek}
\usepackage{subcaption}         
\usepackage{siunitx}      
\usepackage{lscape}             
\usepackage{tabularx}           
\usepackage{placeins}           
\usepackage{orcidlink}
\usepackage{natbib}
\usepackage{ulem}
\usepackage{CJKutf8}
\newcommand{\namecn}[1]{\begin{CJK*}{UTF8}{gbsn}({#1})\end{CJK*}}

\bibpunct{(}{)}{;}{a}{}{,} 

\makeatletter
\let\linenumbers\relax

\makeatother

\begin{document}

   \title{Thermal emission spectra of the ultra-hot Jupiter WASP-33\,b}
   \titlerunning{A\&A,xxx,xxx(2026)}

   \author{Qianyi Zou\namecn{邹乾一}\orcidlink{0000-0001-5469-6443}\inst{1,2,3},
           Meng Zhai\namecn{翟萌}\orcidlink{0000-0003-1207-3787}\inst{1,3},
           Wei Wang\namecn{王炜}\orcidlink{0000-0002-9702-444}\inst{1,3}\thanks{E-mail: wangw@nao.cas.cn},
           Guo Chen\namecn{陈果}\orcidlink{0000-0003-0740-5433}\inst{4},
           Enric Pallé\orcidlink{0000-0003-0987-1593}\inst{2,5},
           Fei Yan\namecn{严飞}\orcidlink{0000-0001-9585-9034}\inst{6,7},
           Huan-Yu Teng\namecn{滕环宇}\orcidlink{0000-0003-3860-6297}\inst{1},
           Qinglin Ouyang\namecn{欧阳青林}\orcidlink{0009-0001-1490-2991}\inst{6},
           Yaqing Shi\namecn{石亚卿}\inst{8},
           Li Zhou\namecn{周丽}\orcidlink{0000-0003-2391-0093}\inst{1},
           Zewen Jiang\namecn{姜泽文}\orcidlink{0000-0002-0486-5007}\inst{9},
           Yujuan Liu\namecn{刘玉娟}\orcidlink{0009-0008-3430-1027}\inst{1,3},
           Thomas Henning\inst{10},
           Nicolas Crouzet\inst{11},
           Gang Zhao\namecn{赵刚}\inst{1,3}
        }
    \authorrunning{Q.Y.Zou et al.}

   \institute{National Astronomical Observatories, Chinese Academy of Sciences, Beijing, 100101, PR China\\
   \and Instituto de Astrofísica de Canarias, 38205 La Laguna, Tenerife, Spain\\
   \and School of Astronomy and Space Science, University of Chinese Academy of Sciences, Beijing, 100049, PR China\\
   \and CAS Key Laboratory of Planetary Science, Purple Mountain Observatory, Chinese Academy of Sciences, Nanjing, 210023, PR China\\
   \and Departamento de Astrofísica, Universidad de La Laguna, 38206 La Laguna, Tenerife, Spain\\
   \and Department of Astronomy, University of Science and Technology of China, Hefei 230026, PR China\\
   \and Deep Space Exploration Laboratory, Hefei, Anhui 230026, China\\
   \and Department of Science Research, Beijing Planetarium, Beijing, 100044, PR China\\
   \and State Key Laboratory of Particle Astrophysics, Institute of High Energy Physics, Chinese Academy of Sciences, Beijing 100049, China\\
   \and Max Planck Institute for Astronomy, Max Planck Institute for Astronomy, K\"{o}nigstuhl 17, D-69117 Heidelberg, Germany\\
   \and Kapteyn Astronomical Institute, University of Groningen, P.O. Box 800, 9700 AV Groningen, The Netherlands\\
   }

   \date{Received 10 11, 2025}
 
\abstract{Observations of exoplanetary atmospheres provide critical insights into their chemical composition, formation, and evolution history. Ultra-hot Jupiters serve as excellent targets for atmospheric characterization; studies of these planets may yield key understanding of gas giants' formation and evolution history. We present a thermal emission study of WASP-33\,b's dayside atmosphere, based on two secondary eclipse observations with CFHT/WIRCam in two specific narrow band filters, namely the CO and CH4$_{\rm on}$ filters, and archival data with HST/WFC3 and Spitzer. Stellar pulsations of the host star induce some quasi-periodic photometric variations, particularly in the CH4$_{\rm on}$ band, which are modeled and corrected in the high-precision differential light curves. An eclipse depth of $1565.2^{+228.6}_{-237.5}$\,ppm and $914.3^{+56.1}_{-57.0}$\,ppm is determined  for the CO and CH4$_{\rm on}$ bands, respectively. Combined with HST/WFC3 and Spitzer data, our joint retrieval of WASP-33 b's dayside atmosphere reveals a high metallicity ([Fe/H] $= 1.52^{+0.35}_{-0.52}$), high C/O ratio (C/O $= 0.78^{+0.03}_{-0.04}$), and a thermal inversion layer, suggesting a formation history involving metal-rich gas accretion. We confirm the presence of the molecules H$_{2}$O, H$^{-}$ and CO, and report a tentative detection of TiO in the dayside atmosphere of WASP-33\,b. Future higher precision observations with JWST may provide better understanding of constraints on the chemical abundances of oxygen and refractory element abundances to better constrain WASP-33\,b's formation and evolutionary pathway.}


\maketitle

\section{Introduction}

\label{sec:intro}

Characterizations of exoplanetary atmospheres are vital for addressing some key questions in the field of exoplanets, such as the formation and evolution of planets, the properties and evolution of planet atmospheres, and the detection of possible biosignatures. Over the past two decades, atmospheric studies have predominantly focused on transiting close-in gas giants, particularly the hot Jupiters (HJs) and ultra-hot Jupiters (UHJs), due to their relatively strong atmospheric signals. Joint analysis based on data across broad wavelengths and/or multiple resolutions enables constraints on atmospheric temperature-pressure profiles, chemical composition, dynamics, and escape processes.

Ultra-hot Jupiters, which have high equilibrium temperatures $T_{\rm eq}$>2200\,K~\citep{Parmentier2018, stangret2022, Tan2024} and highly inflated atmospheres, represent a distinct population of planets and are the most favorable targets for time-resolved spectroscopic observations such as transmission and emission spectroscopy~\citep{snellen2025}. Recent abundance retrievals for some HJs and UHJs, for example WASP-121\,b~\citep{smith2024,evans2025} and WASP-18\,b~\citep{sheppard2017}, seem to be inconsistent with predictions for this population, particularly the expected inverse correlation between metallicity and the carbon-to-oxygen (C/O) ratio~\citep{Espinoza2017, Cridland2019}. Although mechanisms such as pebble drift~\citep{booth2017} could potentially explain these abundances, the observed discrepancies invoke the demand for more observation input to test and refine current theories.

Moreover, several UHJs show significant spin-orbit misalignments, including MASCARA-5\,b~\citep{stangret2021}, KELT-18\,b~\citep{rubenzahl2024}, TOI-1518\,b~\citep{cabot2021}, KELT-9\,b~\citep{gaudi2017}, WASP-121\,b~\citep{delrez2016}, and WASP-33\,b ~\citep{watanabe2022}, indicating that UHJs experience dynamically extreme evolutionary histories, such as the eccentric Lidov-Kozai effect~\citep{naoz2011}. Atmospheric studies of these systems may therefore help constrain when and where planetary migration occurred.

UHJs experience stellar irradiation 10-100 times stronger than classic HJs, and 2-6 orders of magnitude stronger than the warm or cool planets. The intense UV flux drives high temperatures that produce thermal inversions in their upper atmospheres~\citep{baxter2020}. Within and above the temperature inversion layer on the dayside, most molecules such as H$_{2}$O, TiO, and VO undergo thermal dissociation, while CO can still remain relatively abundant~\citep{madhusudhan2012,moses2012,drummond2019}. Therefore, atmospheres become dominated by atomic and ionic species (e.g., \ion{Fe}{i/ii}, \ion{Mg}{i}, \ion{Ca}{ii}, \ion{Na}{i}, \ion{Ti}{i}, and \ion{V}{i}).

Observations from ground-based high-resolution spectroscopy (HRS) and space-based low to medium-resolution spectroscopy support the transition from molecular to atomic and ionic species in UHJs, reflecting extensive thermal dissociation at high temperatures, although nightside cold-trapping and rainout at the terminator that have been observed \citep{gandhi2023, hoeijmakers2024} may also play a role. Some UHJs exhibit extended envelopes or tails as revealed from transit or near-transit observations, suggesting extreme mass loss~\citep{yan2018, cabot2020, yan2021}. These observational atmospheric properties are highly valuable for understanding the composition of the UHJs atmosphere and refining 3D atmospheric modeling with general circulation models (GCMs).

The transiting UHJ, WASP-33\,b, discovered by \citet{Christian2006} in the SuperWASP project \citep{Pollacco2006}, has an orbital period of $\sim$1.22 days. The follow-up observation measurements yielded a mass of 2.1\,$M_{\rm J}$ and a radius of 1.6\,$R_{\rm J}$~\citep{Chakrabarty2019}. Notably, its host star WASP-33 (HD\,15082) is a bright, rapidly rotating A5 $\delta$\, Scuti star \citep{grenier1999,2003Cutri}, making it a peculiar exoplanet. Subsequent Rossiter-McLaughlin effect (RME) observations~\citep{Johnson2015,stephan2022,watanabe2022} revealed a near-polar orbit, showing strong resemblance to two other UHJs around A-type stars, i.e., KELT-9\,b~\citep{gaudi2017} and WASP-189\,b~\citep{Anderson2018,Lendl2020}. This configuration suggests an intense migration history, potentially involving mechanisms such as eccentric Lidov-Kozai effects~\citep{naoz2011}. Interestingly, \cite{mugrauer2019} pointed out that WASP-33 is at least a binary or even a hierarchical triple star system, including a close-in M-dwarf companion (WASP-33B; $\rho\sim1.9\arcsec$, $\Delta\,K_{\rm s} = 6.11\pm0.02$; \citealt{moya2011, wollert2015, ngo2016}) and a wide-separation G-dwarf companion~\citep[WASP-33C, $\rho\sim49.0\arcsec$;][]{mugrauer2019}, invoking the possibility of dynamic perturbation on the planet in the past and/or still ongoing.
    
Multi-resolution studies in a broad wavelength range have been conducted extensively for the characterization of WASP-33\,b. Early photometric eclipse measurements at 0.91\,$\upmu$m, the K$_{\rm s}$ band, and the Spitzer/IRAC broadband yielded notably high dayside brightness temperatures ($\ge3300\,$K), suggesting a potential thermal inversion supported or a high C/O ratio~\citep{smith2011, deming2012, de2013}. The first eclipse spectrum, obtained by \citet{haynes2015} using HST/WFC3, showed excess flux at short wavelengths that was attributed to TiO emission. However, a subsequent reanalysis of these data by \cite{changeat2022} and HRS has questioned the presence of TiO ~\citep{Nugroho2017, cont2021, Herman2020, Yang2024a}. Recent optical and Near-infrared (NIR) HRS detected atomic species, including \ion{Fe}{i}~\citep{Nugroho2020}, \ion{Si}{i}~\citep{cont2022b}, \ion{V}{i} and \ion{Ti}{i} \citep{cont2022},  molecules, including CO~\citep{van2023, Yan2022}, OH~\citep{Nugroho2021}, and a weak H$\rm _{2}$O signal~\citep{Nugroho2021}, which reveal thermal inversion and strong dissociation on the dayside. Transmission spectroscopy detected absorptions from \ion{Ca}{ii} ~\citep{Yang2024a} and AlO~\citep{von2019}, and probed an extended envelope with an escape rate of $\sim 10^{12}\,{\rm g\,s}^{-1}$ by measuring additional absorption of Balmer lines. Notably, \citet{Yang2024b} identified a possible contribution of nightside H$\rm _{2}$O emission to transmission spectra, suggesting the presence of nightside water, and that WASP‑33\,b may exhibit nonisothermal temperature structures on its nightside. Phase-curve observations reveal inefficient day-night heat redistribution~\citep{zhang2018, Herman2022, dang2024} and CO absorption on the nightside, making WASP-33\,b the second exoplanet with confirmed nightside molecules \citep{Mraz2024}. However, most constraints remain limited to species detection, with few atmospheric retrievals conducted~\citep{haynes2015, changeat2022, finnerty2023}. These retrievals consistently indicate a dayside thermal inversion, as predicted by \citet{molliere2015} for atmospheres with C/O $\approx$ 1 and Teff $\gtrsim$ 1500 K, yet they differ in molecular abundances and profile details, highlighting atmospheric complexity that requires further observational constraints.

    \begin{figure}
        \centering
        \includegraphics[width=8cm]{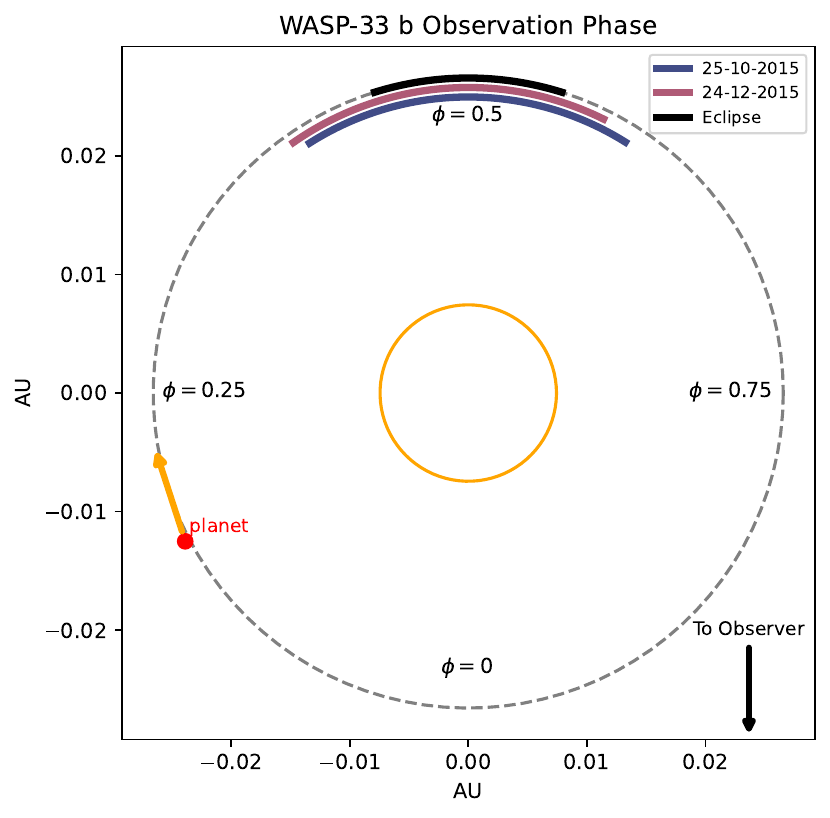}
        \caption{Orbit phases of our two observations. Specifically, Night 1 covered phase 0.410-0.589 (October 25, 2015), Night 2 covered phase 0.403-0.573 (December 24, 2015).}
        \label{fig:phase_coverage}
   \end{figure}

In this work, we report two ground-based high-precision eclipse observations of ultra-hot Jupiter WASP-33\,b obtained with the Canada-France-Hawaii Telescope (CFHT) in 2015 using the CH4$_{\rm on}$ filter and our customized CO filter. Section \ref{sec:obs} details the observational setup, followed by a brief description of data reduction and analysis in Sect. \ref{sec:data}. Section \ref{sec:retrieval} describes our atmospheric retrieval methodology and results, followed by the summary in Sect. \ref{sec:conclusion}.

\section{Observations}
\label{sec:obs}

   \begin{figure}
   \centering
    \includegraphics[width=4cm]{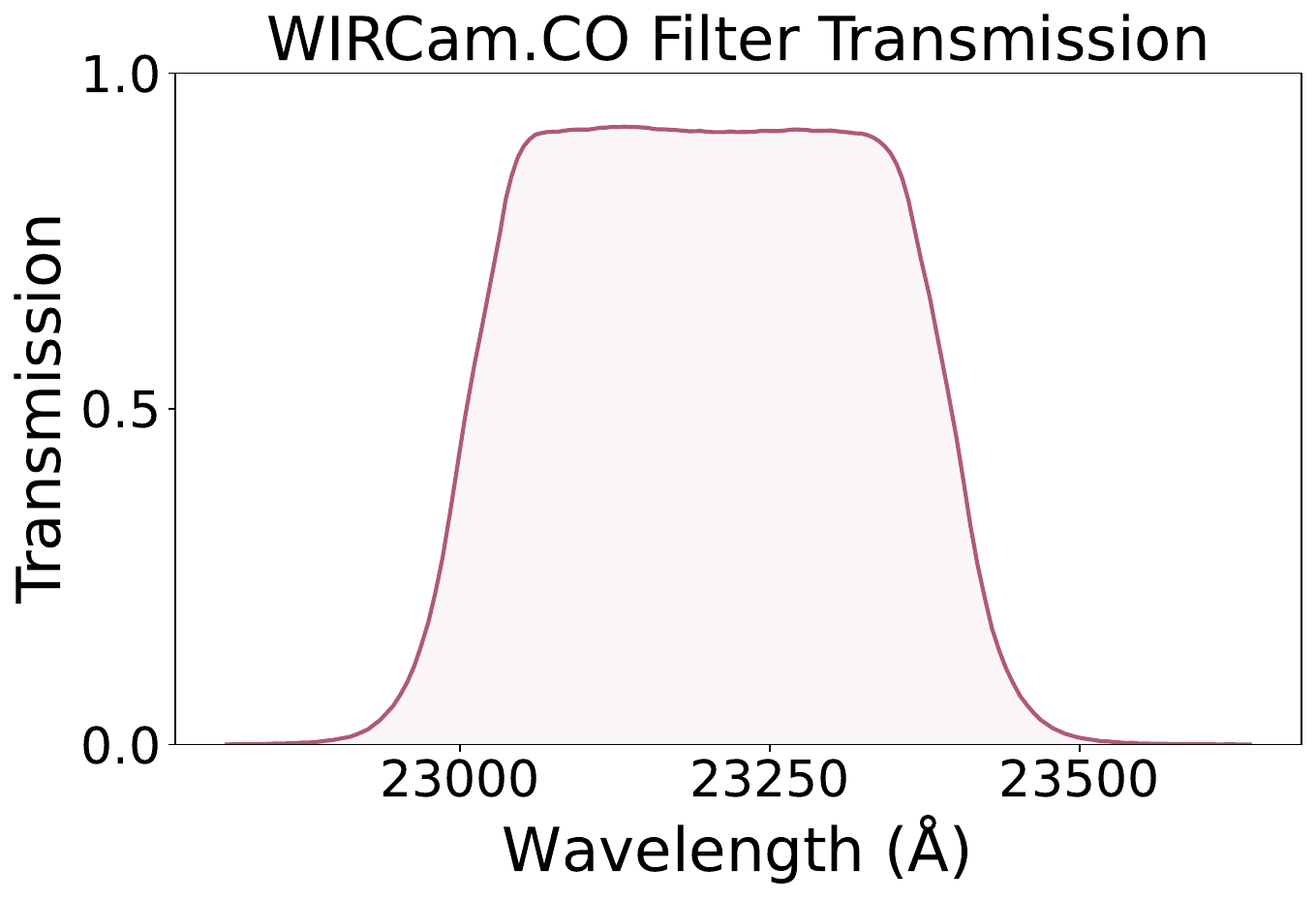}
    \includegraphics[width=4cm]{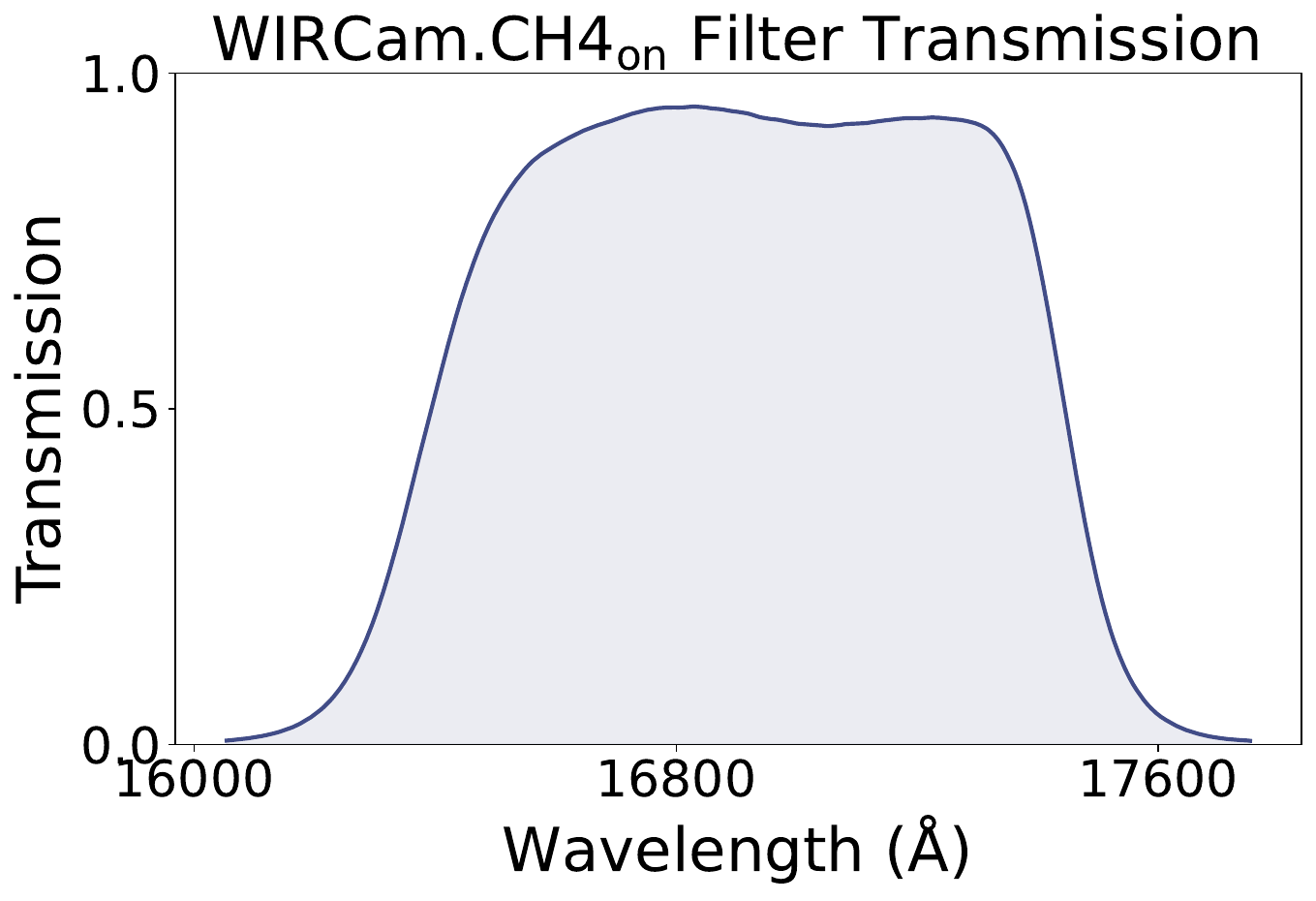}
      \caption{Transmission curves of the CO and CH4$_{\rm on}$ filters used in this work. }
         \label{fig:filter}
   \end{figure}
    
We observed two secondary eclipses of WASP-33\,b (Program 15AS007; PI: Wei Wang) using the Wide-field InfraRed Camera (WIRCam)\footnote{https://www.cfht.hawaii.edu/Instruments/Imaging/WIRCam/} on the CFHT. WIRCam features a 21 arcmin $\times$ 21 arcmin field of view that enables differential photometry with multiple reference stars. Observations occurred on October 25, 2015 in the CO filter (5.2 hours duration) and on 24 December, 2015 in the CH4$_{\rm on}$ filter (6 hours duration). As Fig.~\ref{fig:phase_coverage} shows, both of them fully covered the secondary eclipse along with $\sim$ 2.5 hours of out-of-eclipse baseline. We acquired 696 exposures (12s each) in CO and 996 exposures (3.5s each) in CH4$_{\rm on}$. In order to achieve differential photometric precision to 10$^{-4}$, the \texttt{staring mode}\citep{Devost2010} was used in both observations, following previous successful attempts~\citep[e.g.,][]{croll2010, wangwei2013, chenguo2014, shi2023}. The \texttt{staring mode} aims to maintain the telescope's pointing stability to a few pixels, thereby minimizing the impact of intrapixel and interpixel variations of the detectors on photometric precision. However, our data show that the target star exhibits large movement with maximum shifts of $\sim$15 pixels in the CO filter. This drift likely results from the guiding system misidentifying a cluster of hot pixels on Detector $\#$77 as a stellar point during observations. The misidentification also leads to World Coordinate System (WCS) interpretation failures for the affected frames, rendering them unusable. In contrast, target movement remains within 5 pixels in the CH4$_{\rm on}$ filter. Figure~\ref{fig:filter} shows the filter information of the CFHT/WIRCam's CO and CH4$_{\rm on}$ filters, whose $\lambda_{mean}$ are 2.32\,$\upmu$m and 1.69\,$\upmu$m. The CO filter is customized to be sensitive to CO features at about 2.3\,$\upmu$m, while the CH4$_{\rm on}$ filter is sensitive to both CH4$_{\rm on}$ and H$^{-}$. We use these two filters to assess the presence and abundances of CO, CH4$_{\rm on}$, and H$^{-}$, and thus to provide constraints on the C/O ratio and metallicity.

\section{Data Analysis}
\label{sec:data}

\subsection{Raw data reduction and correction}

Observations obtained with CFHT/WIRCam utilized the sample-up-the-ramp (nondestructive readout) mode~\citep{finger2008}, with each integration sequence consisting of 12 reads. Our analysis reveals that the initial read of each sequence on Detector \#52 consistently exhibited elevated counts, producing photometric values $1.002-1.004$ times higher than the median of other reads, which cannot be reliably corrected. Hence, we excluded the first read of each sequence and only used the remaining 11 reads in this work.

\begin{figure}
        \centering
        \includegraphics[width=8cm]{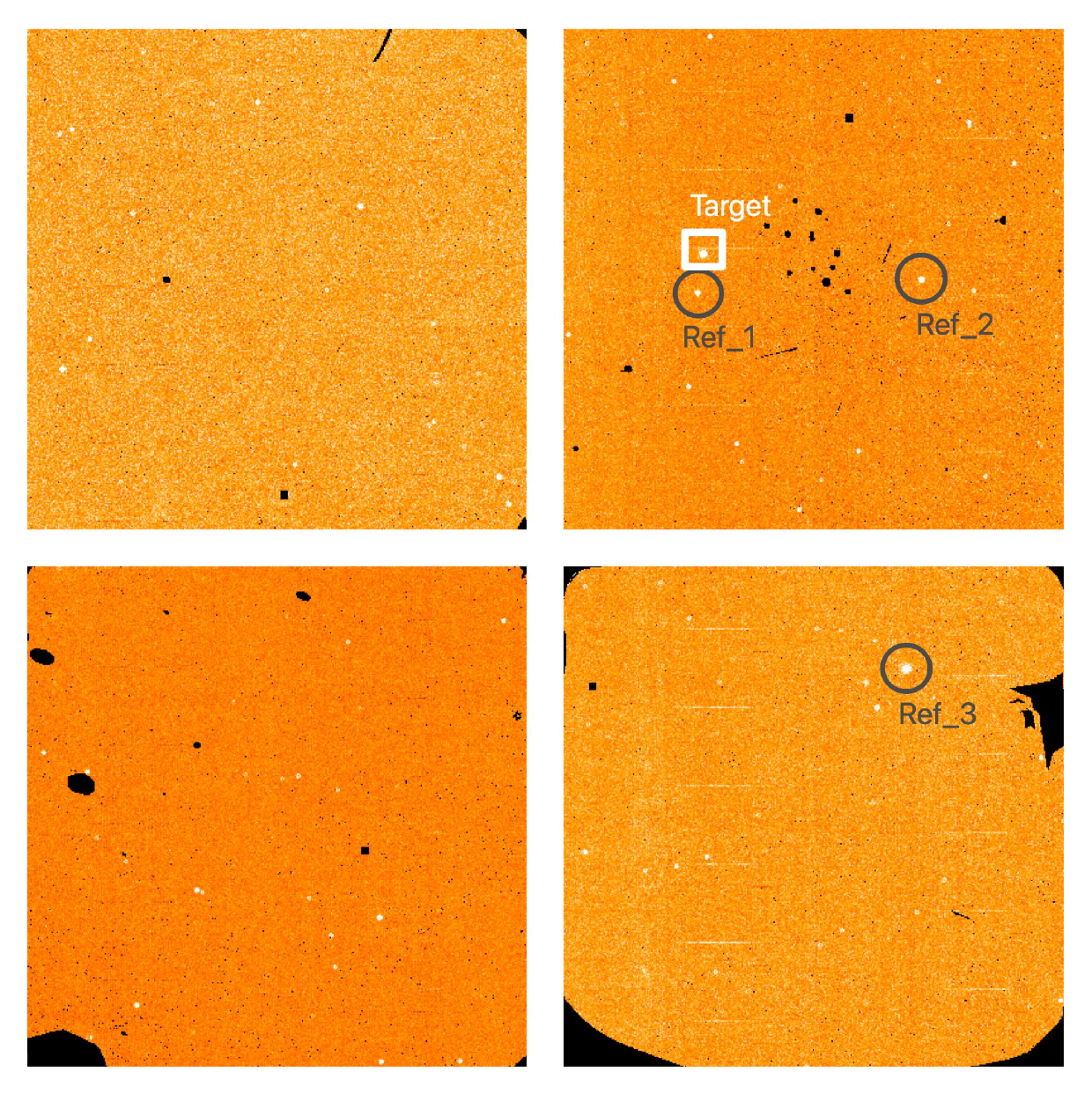}
        \caption{CO filter reduced full-frame WIRCam image of WSAP-33 taken on October 25, 2015. The target star is marked as the white square and the final chosen reference stars are marked as the grey circles.}
        \label{fig:image}
   \end{figure}

Raw data were reduced by the \texttt{`I`iwi} pipeline (version 2.1.200)\footnote{https://www.cfht.hawaii.edu/Instruments/Imaging/WIRCam\newline /IiwiVersion2Doc.html}, including flagging saturated pixels, nonlinearity correction, reference pixels subtraction, dark subtraction, flat fielding, bad pixels masking, and guide window masking. Technically, the \texttt{`I`iwi} pipeline should assign bad pixels and saturated pixels' count a value of Not-a-Number (NaN). However, we find that several pixels within the stellar profile are misclassified as saturated pixels by \texttt{`I`iwi}, resulting in their assignment of NaN values. This misclassification further exacerbates the challenges associated with our data reduction and analysis. Figure~\ref{fig:image} shows a full-frame CO filter's reduced image illustrating our observation field of view (FOV), in which the target star is marked as the red square, and the final chosen reference stars are marked as the blue circles.

As shown in Fig.~\ref{fig:image}, the WIRCam detectors contain a large number of bad pixels. While reference stars were preferentially selected to avoid regions with clustered bad pixels, there are still some isolated defective pixels occasionally remaining within photometric apertures of some reference stars in some exposures, leading to strong systematic in the yielded lightcurves and thus preventing a high-precision differential photometry. To mitigate this effect, we used the \texttt{Background2D} module of the \texttt{Photutils} package~\citep{photutils2024} for spatial interpolation of NaN values, following the methodology outlined by \citet{Ding2025}. This technique can estimate the data values for the affected pixels, demonstrating reliable performance in handling such detector imperfections. 

    \begin{figure}
        \centering
        \includegraphics[width=8cm]{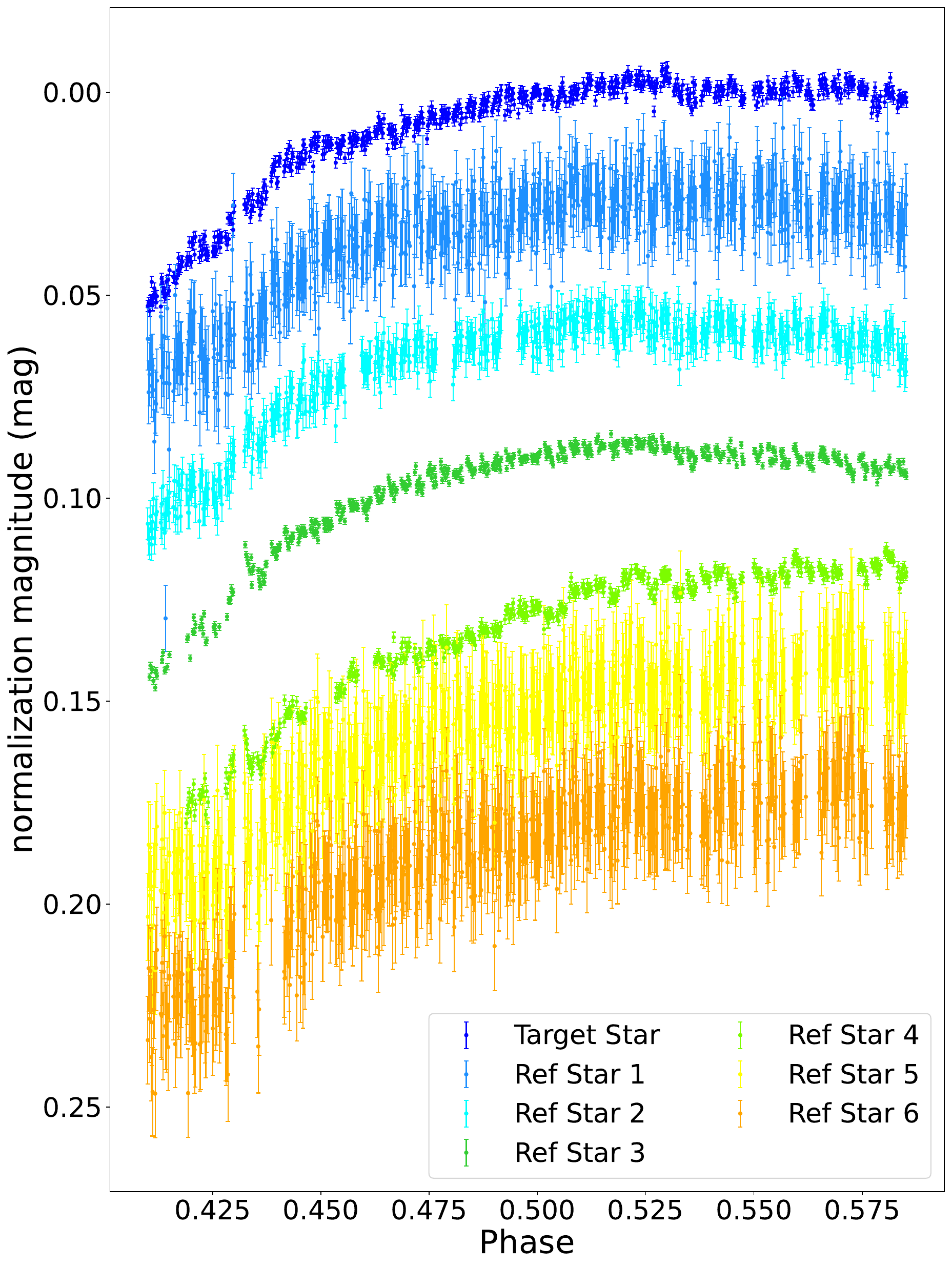}
        \caption{Normalized light curves of target star and the candidate reference stars. }
        \label{fig:lc}
   \end{figure}

\subsection{Photometry and light curve determination}
\label{subsec:lc}

To achieve a photometric precision down to 10$^{-4}$ that is required for the characterization of exoplanet atmospheres, our observations were carried out in defocused mode. With this configuration, the observation efficiency is maximized to gain more photons in the eclipse, and the uncertainties in flat-fielding, intrapixel, and interpixel sensitivity variations are minimized. The disadvantage of such a setup is that the point spread function (PSF) of a star becomes donut-shaped rather than  Gaussian-shaped. In our case, the resulting PSF has a radius of $\sim 10-15$ pixels. We have developed a new pipeline, written in Python, based on and improved on the IDL pipeline used in \cite{shi2023}. Our pipeline employs a parallel processing approach, reducing the photometry and light-curve modeling time for a single band from $\sim$1 day to $\leq$40 minutes.
    
We followed the high-precision differential aperture photometry algorithm first proposed by \citet{everett2001} and developed by \citet{croll2010, wangwei2013, shi2023} for the data reduction, light curve detrending and final light curve determination. We initially selected all stars with comparable brightness ($\Delta$m$<$ 2.5 mag) within the FOV as candidate reference stars, preferentially selecting those located within the same detector where the target star resides. To ensure photometric coherence, we further excluded those candidates exhibiting the out-of-eclipse light curves not similar to those of WASP-33 and other references (e.g., Reference Star 4 in Fig.\ref{fig:lc}), as such discrepancies suggest the star may be intrinsically variable and thus unsuitable as a reference. Unfortunately, WASP-33 is so bright that there are only a few ($\leq 8$) qualified candidates for each filter.

\begin{table}
    \caption{Parameters for the WASP-33\,b system.}  
    \label{table:para}    
                            
\renewcommand\arraystretch{1.4}
\begin{tabular}{l l c }      
    \hline
    \hline    
    \noalign{\smallskip}
        Property & Value & References \\
    \hline
    \noalign{\smallskip}
        \multicolumn{3}{c}{WASP-33} \\
    \hline
       Property & Value & References \\         
    \hline                      
       RA & 02:26:51.06 & [1] \\    
       DEC & +37:33:01.7 & [1] \\
       Spectral Type & A5 & [2] \\
       J$_{\rm band}\,(mag)$ & 7.58 & [3] \\
       K$_{\rm band}\,(mag)$ & 7.47 & [3] \\
       Radius\,(R$_{\sun}$) & 1.602$^{+0.062}_{-0.055}$  & [4] \\
       Mass\,(M$_{\sun}$) & 1.653$^{+0.350}_{-0.187}$  &  [4] \\
       T$_{\rm eff}$\,(K) &  7308$^{+133}_{-109}$ & [4] \\
       Metallicity\,(dex) & 0.1 & [5]\\
    \hline
        \multicolumn{3}{c}{WASP-33 b} \\
    \hline
        Period\,(d) & 1.219870$\pm$1$\times$10$^{-6}$ & [6] \\
        T$_{0}$\,(d) & 2454163.223670$\pm$0.00022 & [6] \\
        a$/$R$_{\rm *}$ & 3.571$\pm$0.010 & [6] \\
        R$_{\rm p}$\,(R$_{\rm J}$) & 1.593$\pm$0.074 & [6] \\
        $i$\,(\si{\degree }) & 86.63$\pm$0.03 & [6] \\
        Mass\,(M$_{\rm J}$) & 2.093$\pm$0.139 & [6] \\
    \hline
    
\end{tabular}
    \tablefoot{[1]\citet{gaia2020},[2]\citet{grenier1999},[3]\citet{2003Cutri},[4]\citet{Stassun2019},[5\citet{zhang2018},][6]\citet{Chakrabarty2019}.}
    
\end{table}

For all the selected candidate reference stars, the raw light curves for a given photometry aperture diameter ($D$) were derived and normalized using the median values derived from a $2-3\sigma$ clipping algorithm. Then, we performed systematic quality screening on the light curves and time-series data points. As exemplified by the CO filter data in Fig.\ref{fig:lc}, the first 1,700s data (phase 0.410-0.426) exhibited degraded quality and were therefore excluded. $Ref4$ candidate displayed obviously divergent photometric trends compared to the target star and other reference stars, and was thus discarded from the reference star list. A similar approach was applied to the CH4$_{\rm on}$ filter data, with slightly modified  criteria: the initial and final 960s data (phase 0.403-0.412 and 0.564-0.573) were removed due to significantly worse data quality. These two excisions somewhat reduced the amount of out-of-eclipse baseline data, but improved the overall data quality and reduced the scatter.

Then, an average reference light curve for a given aperture $D$ and reference star group (RSG), denoted $L_{\rm ref(D,RSG)}$, was obtained by taking the weighted mean of the normalized light curves with weights proportional to the inverse variance of their photometric uncertainties ($w_{i} \propto 1/\sigma_{i}^{2}$); this averaged light curve was then used to divide the target light curve so that most systematics were removed. Finally, a 3$\sigma$ clipping algorithm was applied to remove obvious outliers in the normalized target light curve.
    
The derived target light curves were modeled using \texttt{batman}~\citep{Kreidberg2015}, with the modeling parameters summarized in Table~\ref{table:para}. To provide a reasonable reference for the eclipse depths in the CO and CH4$_{\rm on}$ filters, we built a forward model of WASP-33\,b's dayside atmosphere with \texttt{petitRADTRANS}~\citep{molliere2019} using the retrieval results of \citet{finnerty2023}, and then compared the modeled planetary spectrum to the host star's stellar spectrum; accounting for the filters' bandwidths and transmission profiles with the \texttt{species}~\citep{stolker2020} yielded the reference eclipse depths. Although background-related polynomial baselines of the NIR secondary eclipse light curves have been widely reported\citep{croll2010,wangwei2013,chenguo2014,shi2023}, we did not detect such features in our data. This absence can probably be attributed to the dominant influence of WASP-33's pulsations, which may significantly overshadow  instrumental or background effects.
    
    \begin{figure}
        \centering
        \includegraphics[width=9cm]{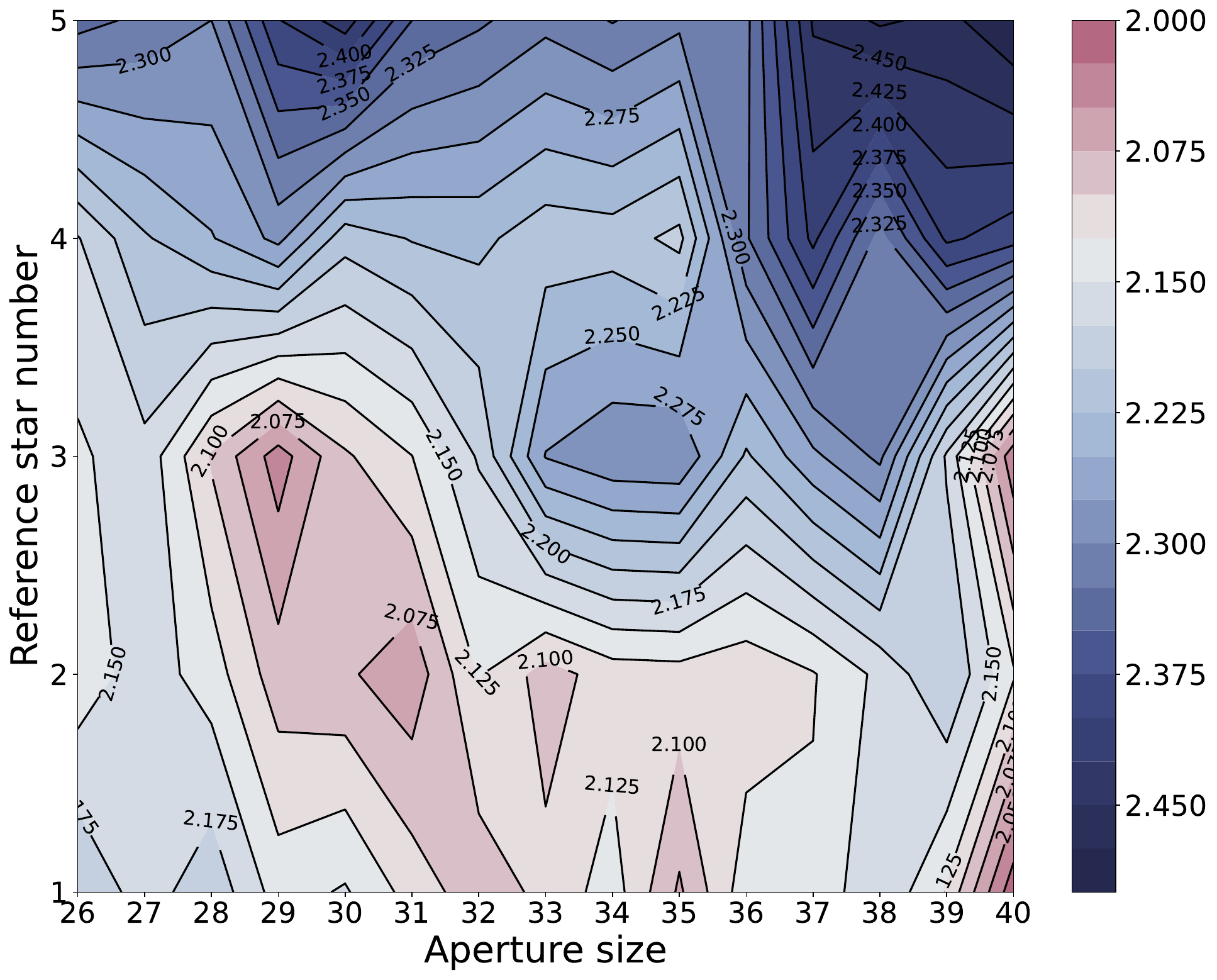}
        \caption{Contour maps of the normalized rms$*\beta^2$ distribution are superimposed on the two-dimensional parameter space defined by $D$ and $N_{\rm RSG}$ in the CO filter data. The pixel scale is 0.306 arcsec $\rm pixel^{-1}$. The minimum value of rms$*\beta^2$ is reached with $D=29$ and $N_{\rm RSG}=3$.}
        \label{fig:rmsbeta_CO}
   \end{figure} 

\subsection{Finding the "best" light curve}
\label{subsec:bestlc}

It is well recognized in our previous work and the literature that for almost every combination of $D$ and RSG, the obtained light curve can be fitted with an eclipse-wise light curve plus various noises, with slightly different eclipse depths, uncertainties and best-fitting residuals. This highlights the need to find the "best" combination of $D$ and RSG, which gives the "best'' light curve that has the smallest fitting residuals with no noticeable hint of overfitting. To do so, we follow the methodology described in \cite{shi2023}, developed from \citet{croll2015}. The key idea of this method is to find a balance between underfitting and overfitting, by combining the two parameters, rms and $\beta$, where rms is the root-mean-square of the residual light curve after subtracting the best-fit model, and $\beta$ is the ratio of the residuals to the Gaussian noise expectation as defined in \cite{winn2008}.The method to derive $\beta$ is described in detail in Sect.~\ref{subsec:lc_fitting}.

When the photometric aperture $D$ encompasses a region where stellar photons dominate the sky background, the residual rms tends to decrease with increasing $D$, gradually approaching the optimal aperture. However, as $D$ continues to grow and the aperture begins to include the outer regions of the stellar profile, the red noise becomes increasingly nonnegligible. This leads to a flattening of the rms curve near its minimum or to a slow upward trend. In such cases, the $\beta$ factor amplifies subtle variations in rms, helping to distinguish and exclude aperture choices with comparable rms values that are affected by correlated noise. Therefore, we use rms and $\beta$ as proxies to achieve a balance between the two competing parameters. In practice, for each set of RSG and $D$, we derive the differential target-to-reference-assembly light curve. The next step is to model all (if possible) systematics (or correlated noise) embedding in the derived light curve, and remove them, to achieve a reliable and high-precision eclipse and the minimum rms and $\beta$. After that, we obtain an rms$*\beta^{2}$ 2D grid, which is smoothed, and thus a minimum value of rms$*\beta^{2}$ can be located.

In the case of WASP-33\,b, an additional approach has to be performed, which is to model and remove the relatively strong stellar pulsation observed in WASP-33. Fortunately, the pattern of the stellar pulsation seems to be irrelevant to the selection of $RSG$ and $D$, and therefore is left to be corrected afterward. 

In Fig.\ref{fig:rmsbeta_CO}, an rms$*\beta^2$ contour map for the CO filter as functions of $D$ and $N_{\rm RSG}$ is shown, where the minimum is achieved at $D=29$\,pixels and $N_{\rm RSG}=3$. We use the same aperture size for the target and reference stars at this stage, but then perform subsequent refinement by individually varying $D$ for each reference star and the target with $\pm 1$ pixel increments, yielding final apertures of $D=29$\,pixels for the science target and $D=$28,39,39\,pixels for the three reference stars. 

\begin{figure}
        \centering
        \includegraphics[width=9cm]{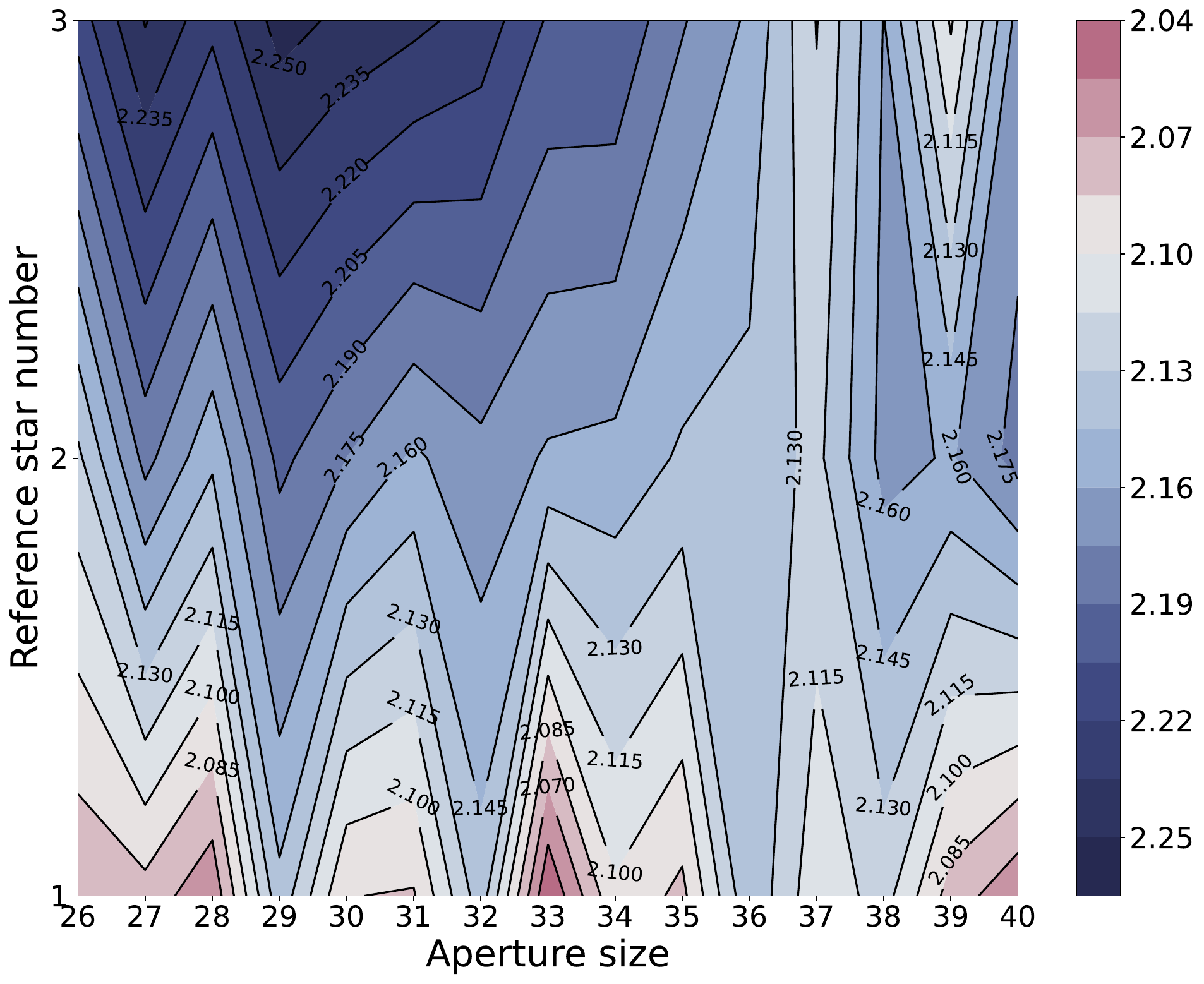}
        \caption{Same as Fig.\ref{fig:rmsbeta_CO} for the CH4$_{\rm on}$ filter. The minimum value of rms$*\beta^2$ is reached with $D=33$ and $N_{RSG}=1$.}
        \label{fig:rmsbeta_CH4}
   \end{figure}

Similar approaches have been applied for the CH4$_{\rm on}$ filter data. However, WASP-33's is so bright in this band that there are only a few usable reference stars (Fig.\ref{fig:rmsbeta_CH4}), resulting in the optimal $N_{\rm RSG}=1$ and $D=33$\,pixels. The final apertures chosen are $D=29$ and 30\,, pixel numbers for the target and reference star, respectively. It is worth noting that in both the CO and CH4$_{\rm on}$ band observations, there seem to be instrumental anomalies in some nondestructive readouts, in which all the 12 sub-integrations show 5$\sigma$ deviations from the nearby readouts; these outliers are flagged and excluded from model fitting.

As noted in Sect.~\ref{sec:intro}, the WASP-33 system may be a triple-star system~\citep{mugrauer2019}. Although the M dwarf companion (WASP-33\,B) can be spatially resolved by a 4m class telescope, it is complicated for our WIRCam observation due to the defocus setup and thus a widespread PSF. Therefore,  partial contamination may occur. We therefore carefully examined the raw images in both filters but found no hint of WASP-33\,B near its expected position, possibly due to the large magnitude difference of $\sim6.11$\,mag in $K_{\rm s}$ and the very short exposure time. Following \citet{shi2023}, we quantified the M dwarf's contaminating signal and found that it contributes only a few parts per million (ppm) to the eclipse depth $-$ far below the measurement uncertainty. The other even fainter companion is quite far away with an angular separation of $\sim49.0\arcsec$ and therefore can be ignored without any risk. We therefore find no evidence for companion-induced contamination in our photometric data and no contamination correction was applied.

    \begin{figure}
        \centering
        \includegraphics[width=9cm]{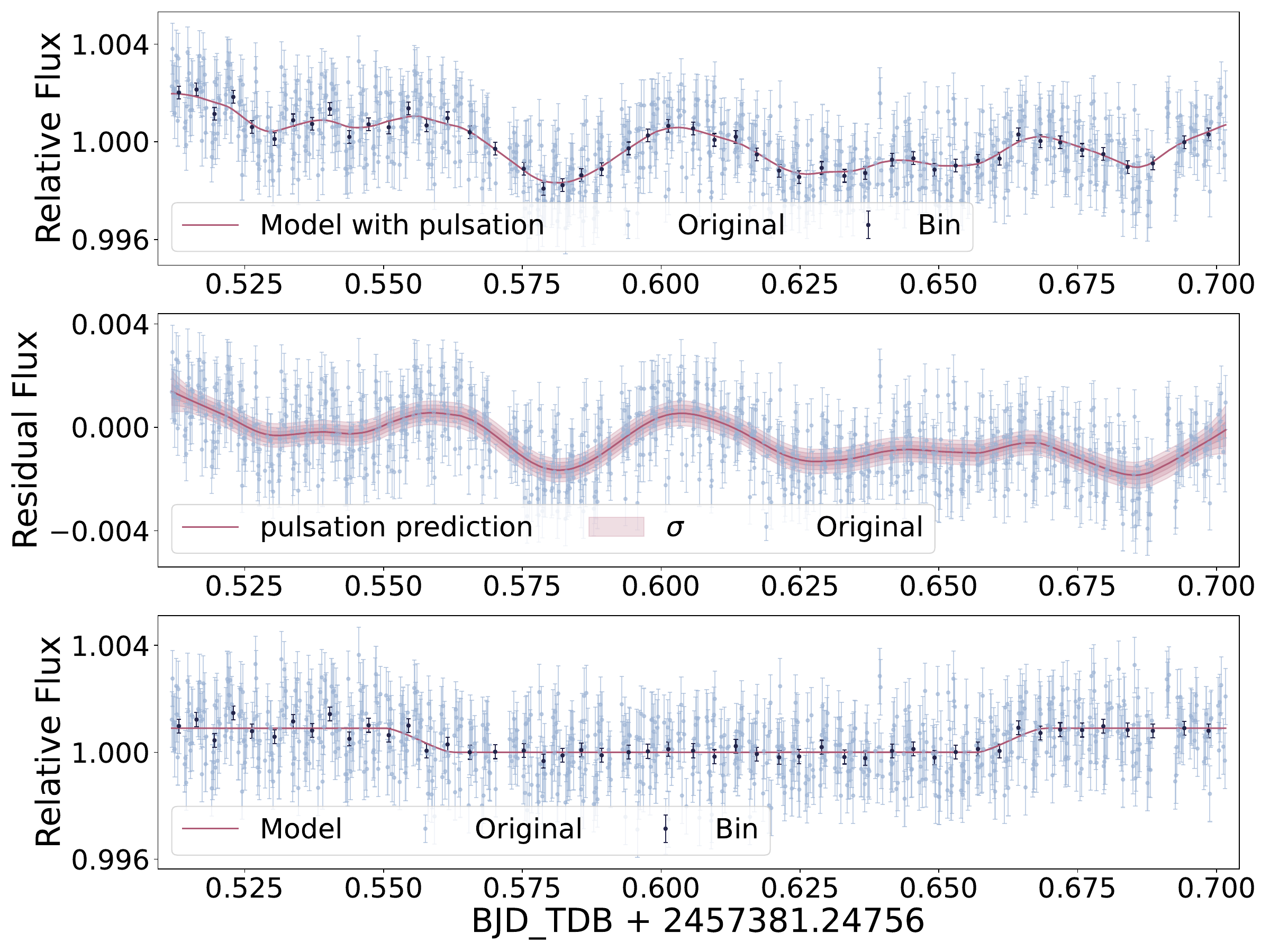}
        \caption{Pulsation noise removal plot for CH4$_{\rm on}$ filter's data. The upper panel illustrates photometric measurements contaminated by pulsational noise, depicted as light points (unbinned) and dark points (binned). A synthesized baseline model (red curve) combines the eclipse signal generated with \texttt{batman} and the pulsational signature predicted by \texttt{celerite2}. The middle and lower panels respectively present the extracted pulsational component and the corrected eclipse light curve after systematic noise removal.}
        \label{fig:pulsation}
   \end{figure}

\subsection{Suppression of stellar pulsation}
    
WASP-33\,b is a rare exoplanet orbiting a variable star. The host star HD\,15082, a $\delta$\,Scuti type star, exhibits a dominant pulsation with a period of $\sim 20$\,days$^{-1}$ and a maximum amplitude of $\sim 1.5$\,mmag~\citep{von2014}. This pulsation may significantly compromise the accuracy of eclipse depth measurements and therefore must be modeled and removed. 

By analyzing the TESS Sector 18 data, \citet{von2020} identified 29 stellar pulsation frequencies with signal-to-noise ratios (S/N)$>$4 in WASP-33. However, \citet{baluev2025} demonstrated that the stellar pulsations may evolve on multi-year timescales after including the TESS Sector 58 data. We tried using the pulsation models derived from \citet{baluev2025} and \citet{von2020} to mitigate the notable photometric variations in the obtained light curves, but the remaining residuals are still nonnegligible and bear some patterns. \citet{deming2012} were able to mitigate the stellar pulsations by combining long-baseline ground-based observations with nearly contemporaneous Spitzer data, enabling the identification of relatively stable pulsation periods that could be modeled and removed from the light curves. However, in our case, the available baseline is considerably shorter, the CFHT and Spitzer observations are widely separated in time, and the pulsation properties of WASP-33 are known to vary. Under these circumstances, applying a similar approach would likely require fitting a large number of additional parameters and may therefore not be practical.

    \begin{figure}[h!]
        \centering
        \includegraphics[width=9cm]{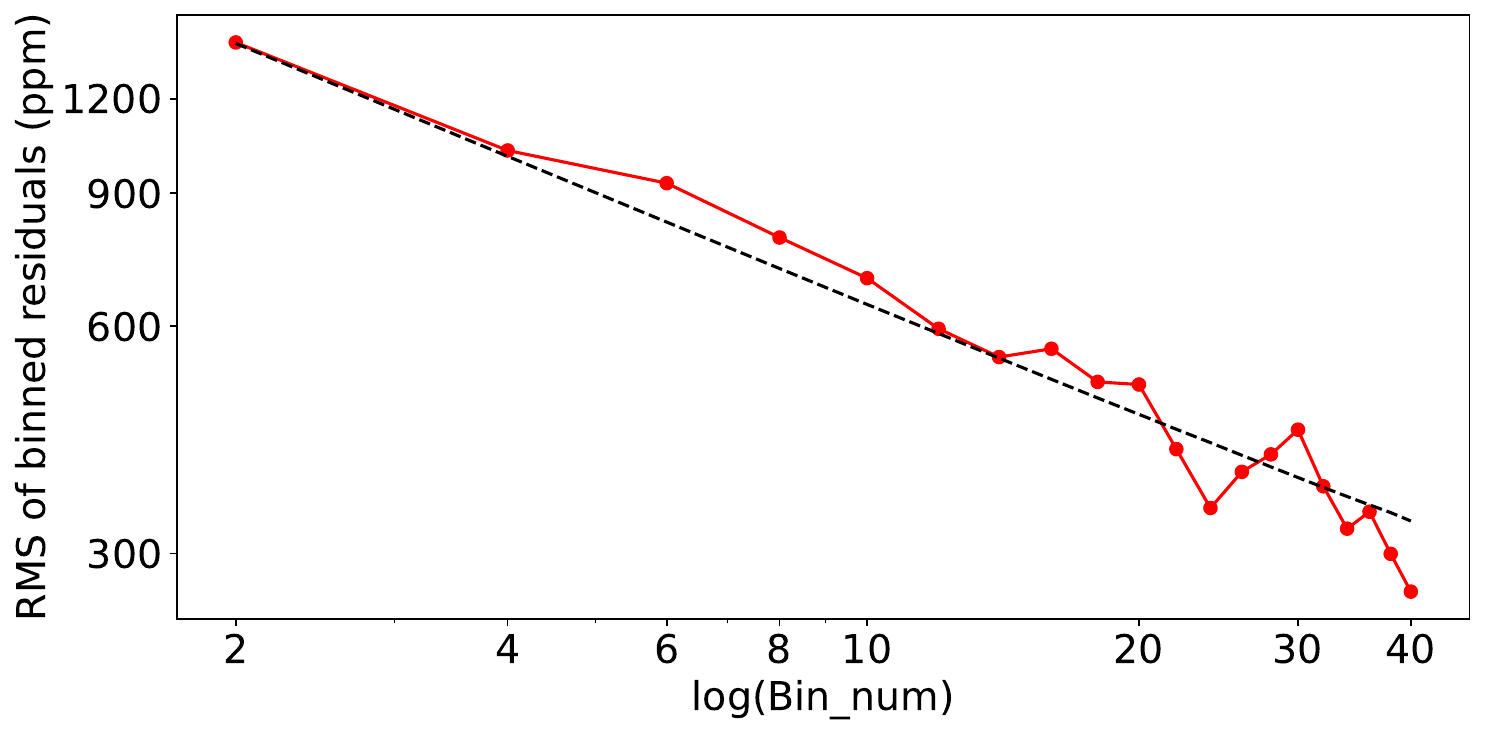}
        \caption{Comparison of data-model residuals and theoretical noise levels across varying bin sizes in CO filter.}
        \label{fig:co_beta}
   \end{figure}

Therefore, we follow Zhang's method \citep{zhang2018} by using a Gaussian process (GP) to fit these pulsations nonparametrically with the public GP code \texttt{celerite2}, \citep{celerite1,celerite2}. The core of this method is that GP can treat pulsations as a combination of various noises whose properties can be described by a parameterized covariance matrix fitted to the data. By removing the requirement to prescribe a mathematical form to the oscillations, this method enables nonstationary oscillatory behavior throughout the observations. The \texttt{celerite} framework constructs the covariance matrix through a kernel function exclusively parameterized by temporal separation $|\tau = t_{i} - t_{j}|$. The kernel used here consists of two radial functions, inspired by a simple harmonic oscillator, augmented with a diagonal component to account for the white noise, which is
    
   \begin{equation}
    \label{eq:k_tau}
      k (\tau) = S_{0}\omega_{0} Q e^{-\frac{\omega_{0}\tau}{2Q}}(cos(\eta\omega_{0}\tau)+\frac{1}{2 \eta Q}sin(\eta\omega_{0}\tau))
\, ,
   \end{equation}
   where $\eta = |1 - (4Q^2)^{-1}|^{\frac{1}{2}}$. 

Two sets of kernels are combined to model the stellar pulsations. In the first kernel, all three parameters ($Q, S_{0}, w_{0}$) are allowed to vary in fit, while in the second kernel, $Q$ is fixed to $1/\sqrt{2}$, and only $S_{0}, w_{0}$ are free parameters. These two kernels represent the oscillatory component and the nonoscillator component that decays rapidly with $\tau$ of stellar pulsations, respectively. In addition, we incorporate the eclipse depth as another free parameter in the fitting process to mitigate potential covariance with the forward model assumptions. For the CH4$_{\rm on}$ band data, the broader bandwidth returns a higher S/N, which enables quite robust identification and systematic removal of pulsation-induced noise, as illustrated in Fig.~\ref{fig:pulsation}. Conversely, the pulsation correction in the CO band is less successful, due to the narrower bandwidth and limited data volume, and thus suppressed pulsation signatures.
   
    \begin{figure}[h!]
        \centering
        \includegraphics[width=9cm]{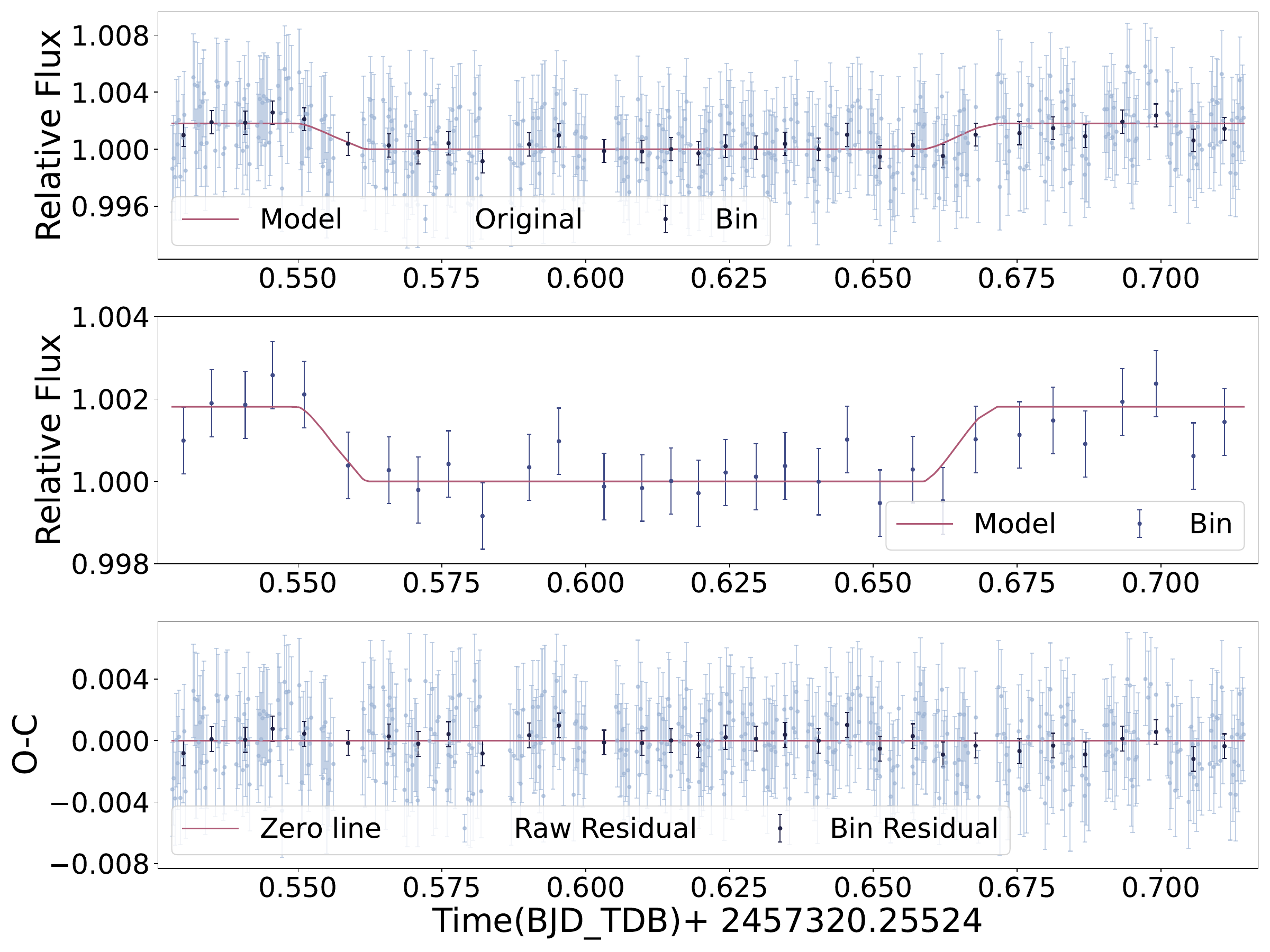}
        \caption{Secondary eclipse of WASP-33\,b observed by CFHT/WIRCam in the CO filter on October 25, 2015. The top panel displays the unbinned (light-shaded points) and binned (dark-shaded points) light curves overlaid with the best-fit secondary eclipse model (solid red line). The middle panel isolates the binned data alongside the optimized eclipse model, while the bottom panel presents the residuals relative to the model fit. The rms of the binned residuals is 544.9 ppm.}
        \label{fig:lc_final_CO}
   \end{figure}

\subsection{Light curve rebinning and fitting}
\label{subsec:lc_fitting}
Until now, the photometric precision of each data point of the obtained light curve is $\sim10^{-3}$, the same order of magnitude as the eclipse depth of a typical HJ. To obtain a robust determination of the eclipse depth, rebinning of data points before light curve fitting is essential. Although the $rms*\beta^2$ method can identify an optimal bin size together with an optimal $L_{ref}(RSG,D)$, this initial binning is on the light curves obtained using identical $D$ for both the science target and the reference stars, and they still remain contaminated by pulsation noise. After finalizing individual $D$ values for every star and removing pulsation contamination, the bin size for each band is reoptimized.
    
As described previously, we employ the $\beta$ parameter \citep{winn2008} to quantify the level of correlated noise. Taking the CO filter data as an example, $\beta$ is calculated by averaging ratios across bin sizes ranging from 2 to 40 data points (with upper limit set to ensure time sampling size to be less than WASP-33\,b's ingress and egress; see Fig.~\ref{fig:co_beta}). The unphysical cases where the scaled-down residuals fall below Gaussian noise expectations, i.e., $\beta<1$, are discarded. After this evaluation, a bin size of 14 is adopted for the CO filter data. A similar procedure yields a bin size of 18 for the CH4$_{\rm on}$ filter data.

    \begin{figure}[h!]
        \centering
        \includegraphics[width=9cm]{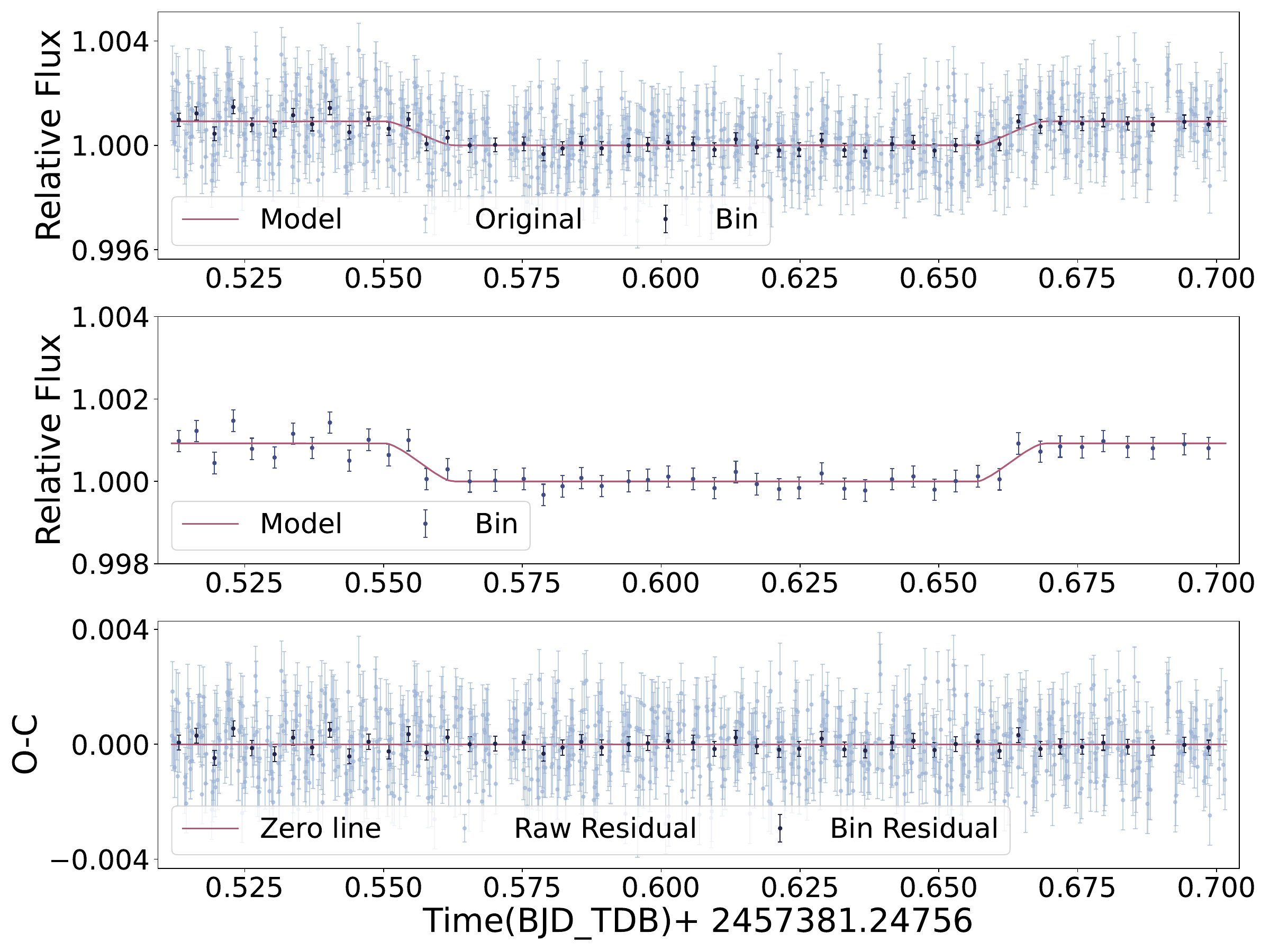}
        \caption{Similar to Fig. 9 but for the secondary eclipse of WASP-33\,b observed by CFHT/WIRCam with the CH4$_{\rm on}$ filter on December 24, 2015. The rms of the binned residuals is 217.1 ppm.}
        \label{fig:lc_final_CH4on}
   \end{figure}

\begin{figure}[h!]
    \centering
    \includegraphics[width=6.8cm]{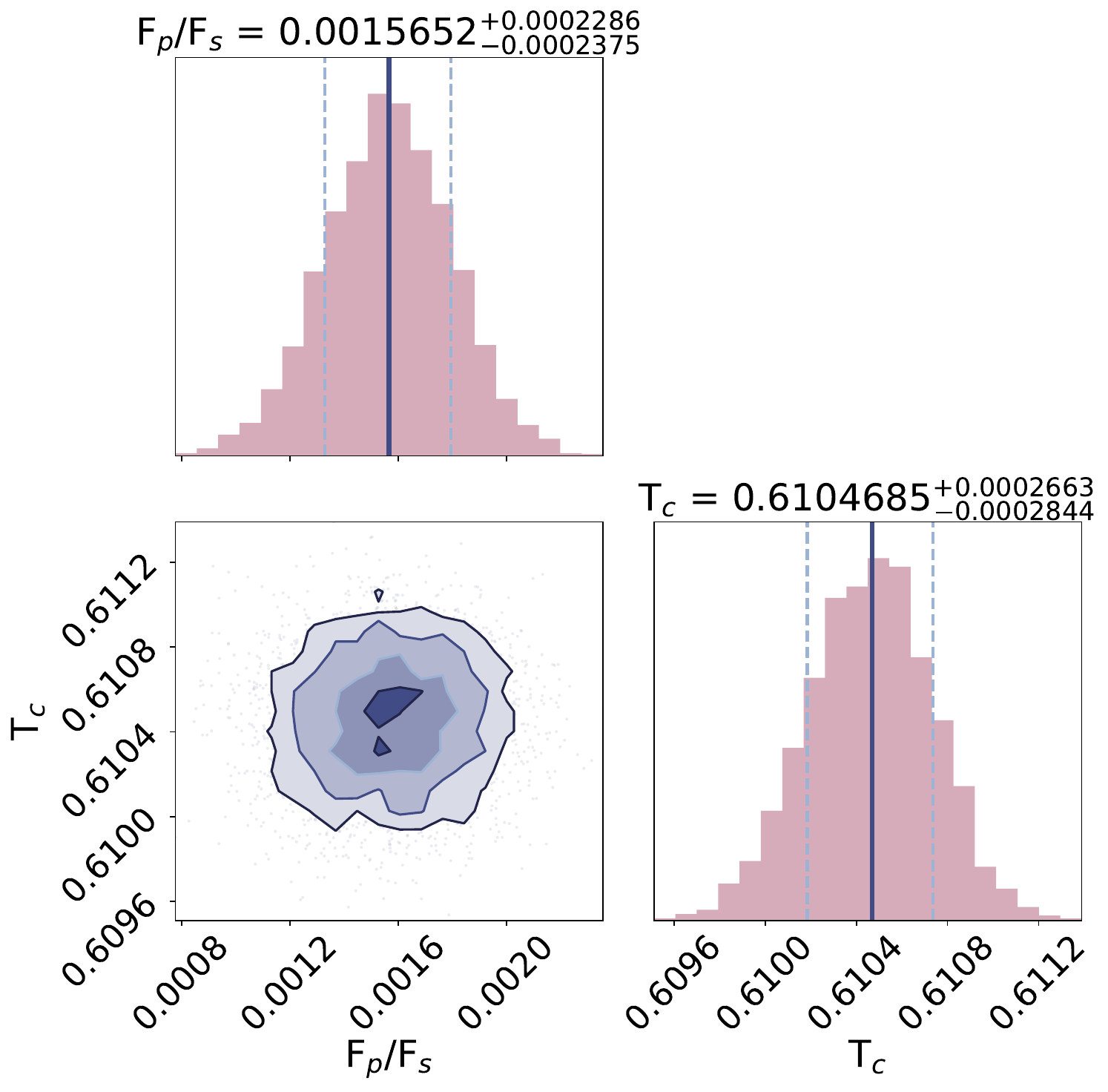}
    \includegraphics[width=6.8cm]{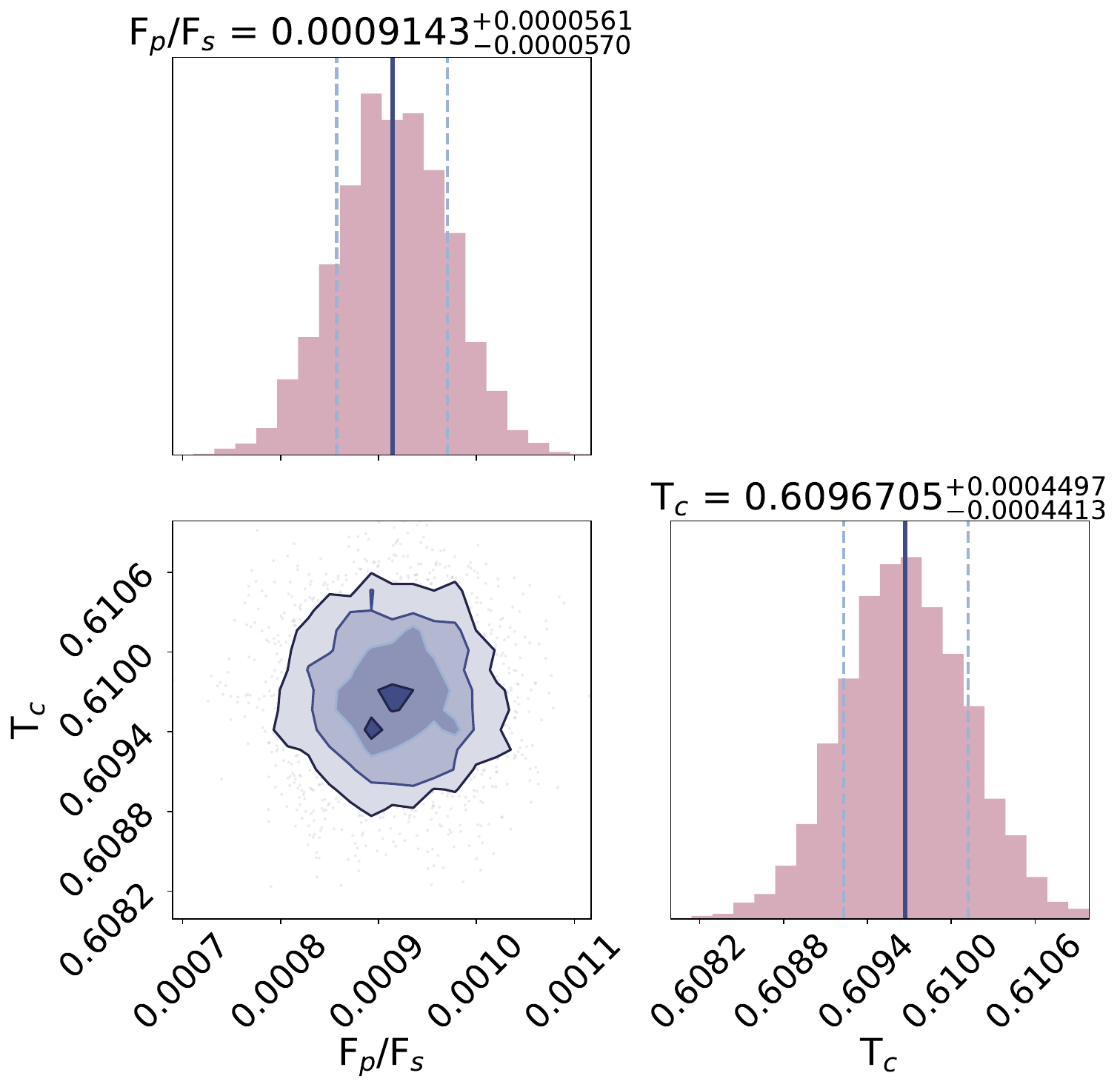}
    \caption{Best-fit parameters of the binned light curve for the CO (\textit{Upper}) and CH4$_{\rm on}$ (\textit{Lower}), with  $T_{\rm mid} = T_{\rm c} - 57319.75524$ and $T_{\rm mid} = T_{\rm c} - 57381.24756 $.}
    \label{fig:corner}
\end{figure}

 The rebinned CO and  CH4$_{\rm on}$ band light curves, as shown in the middle panels of Figs~\ref{fig:lc_final_CO}  and   \ref{fig:lc_final_CH4on}, are used to determine final eclipse depths via Markov chain Monte Carlo (MCMC) analysis using the \texttt{emcee} package \citep{foreman2013}. The MCMC sampling is performed on two parameters, the eclipse depth distributed uniformly, and the central eclipse time $T_{\rm mid}$ distributed normally with sigma being the error of $T_{\rm mid}$,  propagated from the uncertainty quoted in \citet{zhang2018}. Each parameter is sampled with six walkers. An initial 1000-step run refines parameter guesses, followed by a final 6000-step sampling with 12 walkers and a step size of 10$^{-8}$ $>50$ times the best-fit binned data length to avoid possible autocorrelation. After discarding the first 20\% of the chains as burn-in, we confirmed that the acceptance rates remained within 20\%–50\%, consistent with appropriately chosen step sizes. Convergence was further assessed using the Gelman–Rubin statistic ($\hat{R}$), which was found to be close to unity for all parameters. The best-fit eclipse depths in the CO and CH4$_{\rm on}$ bands are $1565.2^{+228.6}_{-237.5}$\,ppm and $914.3^{+56.1}_{-57.0}$\,ppm, respectively (see Fig.\ref{fig:corner}).

\label{sec:retrieval}

\begin{figure*}[ht!]
  \centering
  \includegraphics[width=18cm]{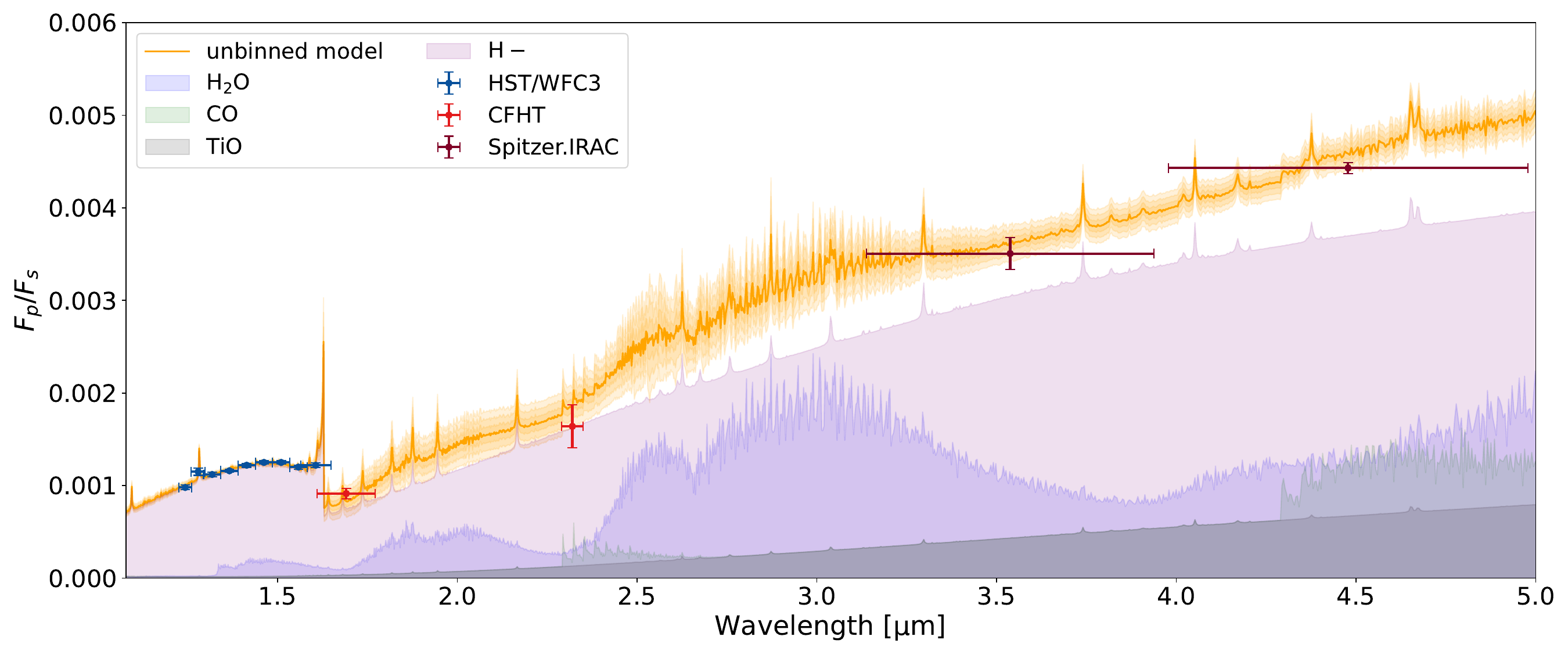}
  \caption{Best-fit FREE model spectrum together with the data points from HST, CFHT, and Spitzer shown in filled dots with error bars and distinct colors. The orange curve and surrounding shaded region represent the best-fit model and its 3$\sigma$ regime. The shaded areas in different colors below the orange curve indicate the reference models containing only H$_{2}$O, CO, TiO, or H$^{-}$, respectively.}
\label{fig:bestfit_free}
\end{figure*}

\begin{figure}[h!]
\centering
\includegraphics[width=6cm]{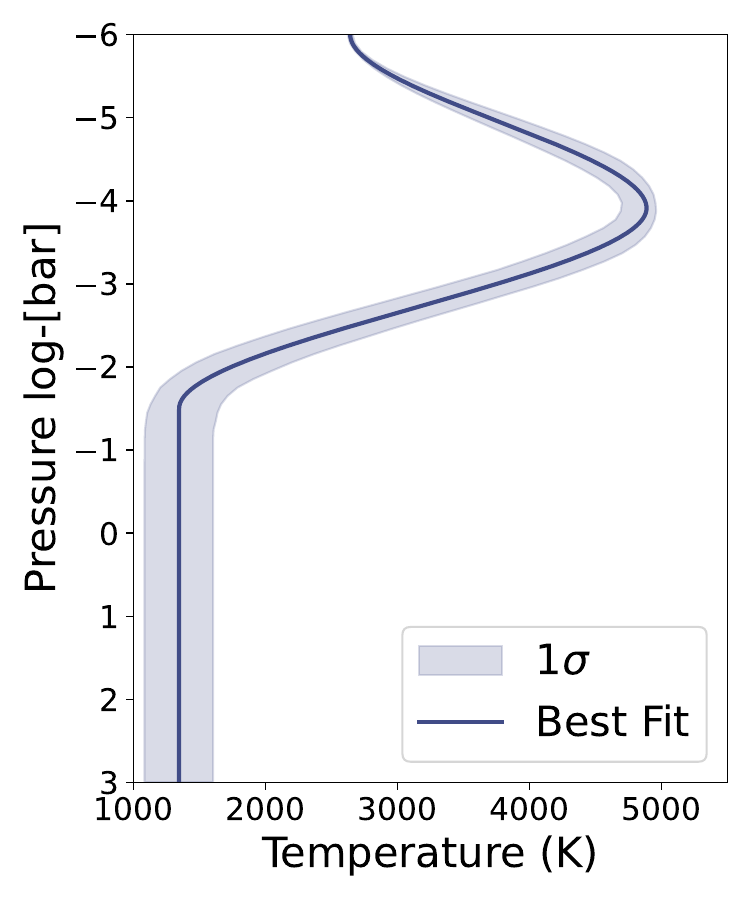}
\caption{Retrieved temperature-pressure (T-P) profile for the FREE model. The solid blue line shows the best-fit solution, and the shaded area denotes the 1$\sigma$ confidence interval derived from the posterior distribution.}
\label{fig:TP_free}
\end{figure}

\subsection{Modeling}
\label{subsec:modelling}

In addition to our newly obtained CFHT/WIRCam CO and CH4$_{\rm on}$ band measurements, for the dayside atmosphere investigation, we collected the HST/WFC3 emission spectrum obtained by \citet{haynes2015} and the Spitzer/IRAC  3.6 and 4.5$\upmu$m photometric eclipse depths obtained by \citet{zhang2018,dang2024}. In total, we have 13 eclipse depth measurements spanning from 1.1$\upmu$m to 4.5$\upmu$m. Note that two HST data points are excluded, as explained in Sec~\ref{subsec:free_retreval}.

We use the \texttt{petitRADTRANS} package \citep{molliere2019,Nasedkin2024} and construct two atmospheric models: a free-chemistry (FREE) model and an equilibrium chemistry (EQ) model. We use the python package \texttt{PyMultiNest} to explore the posterior distribution and calculate Bayesian evidence \citep{buchner2014}, which implements the multimodal nested sampling algorithm based on the MultiNest library \citep{feroz2009}.

\begin{table}
    \caption{Best-fit parameters from the FREE model retrieval on WASP-33\,b.}  
    \label{table:free_model_para}    
\renewcommand\arraystretch{1.4}
\begin{tabular}{l c c c}      
    \hline
    \hline    
    \noalign{\smallskip}
       Parameter & Prior & Posterior distribution & Unit \\        
    \hline                     
       T$_{1}$ & $\mathcal{U}(1000, 3000)$ & 1345$\pm$265 & K \\    
       T$_{2}$ & $\mathcal{U}(3000, 5000)$ & 4887$\pm$110 & K \\
       T$_{3}$ & $\mathcal{U}(1000, 3000)$ & 2641$\pm$20 & K \\
       P$_{1}$ & $\mathcal{U}(-2, 3)$ & $-1.50 \pm 0.24$ & log bar \\
       P$_{2}$ & $\mathcal{U}(-6, -2)$ & $-3.91 \pm 0.22$ & log bar \\
       P$_{3}$ & -6 & ... & log bar \\
       H$^{-}$ & $\mathcal{U}(-10, -1)$ & $-2.131 \pm 0.073$ & dex \\
       H & $\mathcal{U}(-10, -1)$ & $-5.5 \pm 3.0$ & dex \\
       e$^{-}$ & $\mathcal{U}(-10, -1)$ & $-2.3 \pm 0.94$ & dex \\
       H$_{2}$O & $\mathcal{U}(-10, -1)$ & $-6.2 \pm 2.5$ & dex \\
       CO & $\mathcal{U}(-10, -1)$ & $-4.9 \pm 2.9$ & dex \\
       TiO & $\mathcal{U}(-10, -1)$ & $-6.6 \pm 2.1$ & dex \\
       CH$_{4}$ & $\mathcal{U}(-10, -1)$ & $-6.9 n\pm 2.1$ & dex \\
    \hline

\end{tabular}
\end{table}

For the FREE model, we assume a one-dimensional atmospheric structure with a temperature inversion layer spanning a pressure range of $10^{-6}$ to $10^{3}$ bar, divided into 100 uniformly spaced layers in logarithmic space. Previous HR study indicates that the temperature of the upper atmosphere is lower than that of the inversion layer \citep{haynes2015, Nugroho2017, Herman2020, finnerty2023}, which means that the commonly adopted 2-point temperature-pressure (T-P) profile \citep{buchner2014} is inadequate. We therefore introduce a third point (T$_{3}$,P$_{3}$). 
Unlike the 3-point T-P profile proposed by \citet{waldmann2015}, we fix P$_{3}$ at 10$^{-6}$ bar rather than treating it as a free parameter. Based on previous studies on WASP-33\,b's dayside, we include CO \citep{rothman2010}, H$_{2}$O \citep{rothman2010}, CH$_{4}$ \citep{yurchenko2017}, and TiO \citep{mckemmish2019}, which possess prominent spectral features in the wavelength coverage of interests. We have not included OH - a photodissociation product of H$_{2}$O - in our model, due to the lack of OH-sensitive features in the data set we use in this work. The mass fractions of these species are treated as free parameters, while H$_{2}$ and He serve as filler gases to maintain a total mass fraction, with a fixed He/H$_{2}$ mass ratio of 0.305. Contributions from Rayleigh scattering by H$_{2}$/He and continuum absorption from H$_{2}$-H$_{2}$, H$_{2}$-He, and H$^{-}$ are considered, and the inclusion of H$^{-}$ necessitates treating the mass fractions of its associated species, e$^{-}$ and H, as free parameters. Given the high temperatures on the dayside, the atmosphere is likely to be cloud-free, and thus no cloud prescription is included in the retrieval process. For the EQ model, we use the same 3-point T-P profile as described above. The free parameters to be explored are the C/O ratio and the metallicity [Fe/H]. Both retrieval models are initially computed at a spectral resolution ($\lambda/\Delta\lambda$) of 1,000 and subsequently binned to match the observational passbands. They share common fixed parameters, including planetary radius of 1.593\,R$_{\rm J}$, surface gravity of 27.114\,$\rm m/s^{2}$, and orbit semi-major axis of 0.02558\,au.

To identify potential species from the emission spectrum obtained, we first computed the Bayesian evidence \citep{kass1995} for the full model (including all species). We then iteratively excluded one species at a time and recalculated the Bayes factors for each reduced model. Following the criteria proposed by \citet{kass1995}, the strength of evidence is categorized as strong ($|\Delta ln \mathcal{Z} | \geq 5$), moderate ($3 \leq |\Delta ln \mathcal{Z} | < 5$), weak ($1 \leq |\Delta ln \mathcal{Z} | < 3$), or inconclusive ($|\Delta ln \mathcal{Z} | \leq 1$.

\section{Retrieval analysis}
\subsection{FREE model retrieval}
\label{subsec:free_retreval}

The FREE retrieval analysis yields a best-fit spectrum that is largely featureless, with residuals exceeding $3\sigma$ at the blue-end datapoints (1.155 and 1.199\,$\upmu$m). These two points dominate the $\chi^{2}$ value, leading to a generally poor fit.

\citet{haynes2015} attributed the unusually large depths at 1.155 and 1.199\,$\upmu$m to a high abundance of TiO, but subsequent joint retrievals \citep{changeat2022} combining HST and Spitzer data and later HRS studies have contested the presence of dayside TiO \citep{Herman2020,cont2021}. To test this hypothesis, we ran several forward models with temperature inversion layers that included only H, He, and TiO, assuming different TiO abundances, and found that the TiO abundance must be $\geq 10^{-4}$ to account for the depths observed 1.155 and 1.199\,$\upmu$m. However, such a high TiO abundance should lead the eclipse depths of the CH4$_{\rm on}$ and CO bands (especially the latter) to be significantly higher than our measured depths, and TiO should be detected in HR studies \citep{Nugroho2017, Herman2020, cont2021, serindag2021}. 

We therefore argue that the large depths of 1.155 and 1.199\,$\upmu$m cannot be due to a high TiO abundance; rather, they come from their imperfect correction of the pulsation signal. \citet{haynes2015} noticed the pulsation signals in their light curves and modeled the pulsation signals using sine functions. However, such a procedure was later found to only be capable of removing $\sim10$\% of pulsation noise \citep{zhang2018}, thus leaving substantial pulsational residuals in at least some of the HST/WFC3 light curves. Given that WASP-33 is an A-type star, pulsational noise affects the precision of eclipse depth measurements much more significantly at shorter wavelengths, and thus the 1.155 and 1.199\,$\upmu$m depths should still be affected by pulsation noise. To mitigate this and given that the two bluest datapoints are inconsistent with the other datapoints, we excluded the 1.155 and 1.199\,$\upmu$m datapoints in the subsequent retrieval analysis.

In addition to HST measurements, two ground-based secondary-eclipse measurements of WASP-33\,b have been reported in the literature: a 0.9 $\upmu$m datapoint from \citet{smith2011} and a Ks-band measurement from \citet{deming2012}. The 0.9 $\upmu$m measurement is not included in our joint retrieval because the transmission curve of the filter used in that observation is not available, preventing self-consistent modeling within \texttt{petitRADTRANS}. In contrast, Ks-band eclipse measurement can be readily incorporated. We therefore performed additional retrievals including this datapoint and found that the inferred atmospheric parameters remained fully consistent with our nominal results, with all differences well within the 1$\sigma$ uncertainties. This demonstrates that our conclusions are insensitive to the inclusion of the Ks-band measurement; consequently, it is not included in the final retrieval presented here.

\begin{table}
    \caption{Bayesian model comparison of the chemical species considered for the FREE model.}  
    \label{table:BE_free}    
    \renewcommand\arraystretch{1.4}
    \centering                    
\begin{tabular}{l l l }  
    \hline
    \hline
       Species & $\Delta ln(\mathcal{Z})$ & Inferences \\
    \hline                      
       H$_{2}$O & -0.3 & Inconclusive \\    
       CO & 0.1 & Inconclusive \\
       CH$_{4}$ & -0.2 & Inconclusive \\
       $\mathbf{H^{-}}$ & \textbf{71.3} & \textbf{Strong} \\
       TiO & -0.5 & Inconclusive \\
    \hline
\end{tabular}

    \tablefoot{The logarithmic Bayesian evidence, calculated as $ln(\mathcal{Z}_{\text{full}}) - ln(\mathcal{Z}_\text{{no-X}})$, quantifies the statistical preference for excluding a certain species X in the atmospheric model, with uncertainties of 0.7. The species in bold possess strong evidence.}
\end{table}

The best-fit result for the FREE model is shown in Fig.\ref{fig:bestfit_free}, with Figs.\ref{fig:TP_free} \& Table~\ref{table:free_model_para} illustrating the corresponding mass mixing ratios (MMRs) and T-P profile. Table~\ref{table:BE_free} reports the detection significance represented by Bayesian evidence for each chemical species. 
    
In the FREE retrieval, WASP-33\,b's dayside atmosphere exhibits a strong temperature inversion layer and a featureless emission spectrum, resulting in no or very poor constraints on the MMRs for all species considered except H$^{-}$ and e$^{-}$, which have quite high concentrations. The lack of concrete detection of molecules is in line with the low resolution (LR) space-based results from \citet{changeat2022}, but seems to conflict with the HR results \citep{finnerty2023}. The results may be caused by: (1) high temperatures in the emission-contribution layers that thermally dissociate most molecules and thus reduce their detectability; (2) residual pulsation noise in the HST spectra, which could still obscure molecular features, even after correction; and (3) the featureless spectrum forcing the retrieval to fit continuum opacity by the inclusion of a large amount of H$^{-}$.

\begin{figure*}[t!]
\centering
\includegraphics[width=18cm]{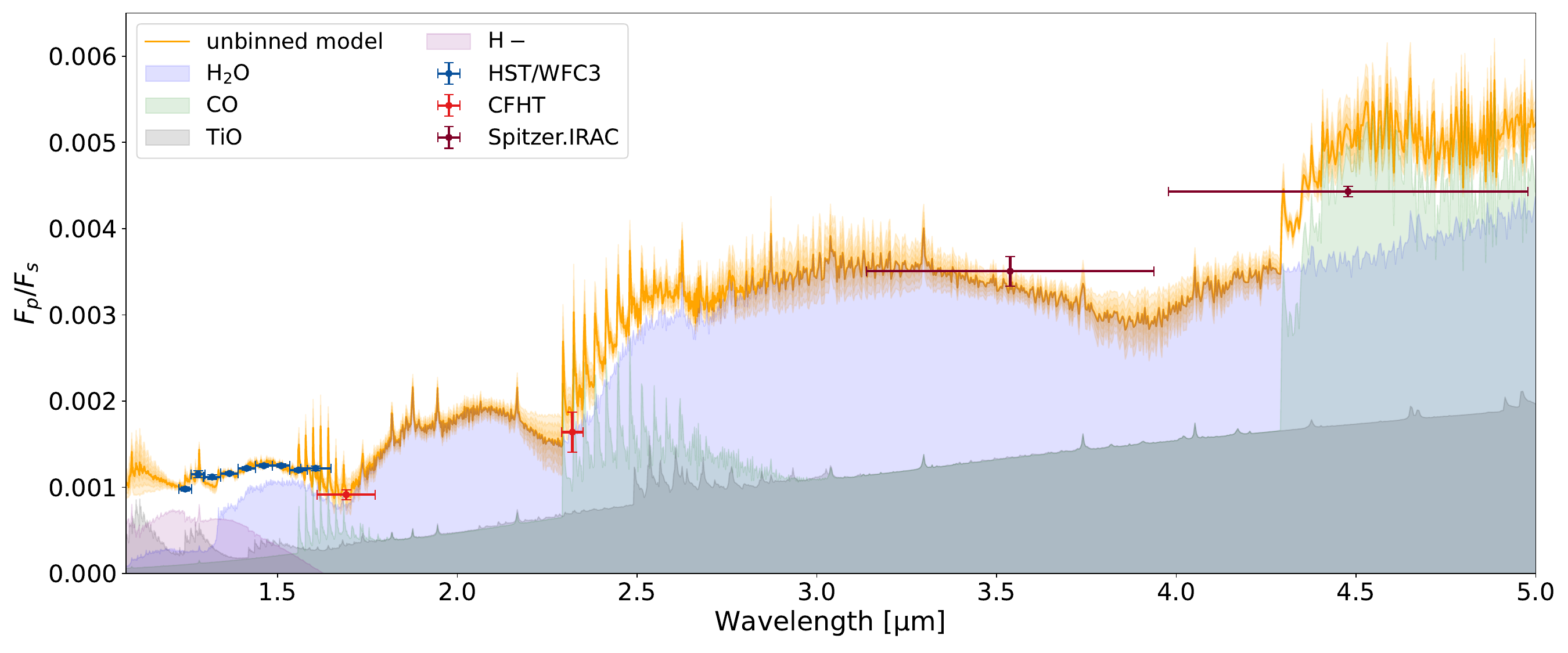}
\caption{Same as Fig.~\ref{fig:bestfit_free}, but for the EQ retrieval. Note that H$_{2}$O, H$^{-}$, and CO all exhibit prominent features in the model spectrum.}
\label{fig:bestfit_eq}
\end{figure*}

\begin{table}
    \caption{Best-fit parameters from the EQ model retrieval on WASP-33\,b.}  
    \label{table:eq_model_para}    

\renewcommand\arraystretch{1.4}
\begin{tabular}{l c c c}      
    \hline
    \hline    
    \noalign{\smallskip}
       Parameter & Prior & Posterior distribution & Unit \\   
    \hline                     
       T$_{1}$ & $\mathcal{U}(1000, 3000)$ & 2034$\pm$338 & K \\    
       T$_{2}$ & $\mathcal{U}(3000, 5000)$ & 4849$\pm$149 & K \\
       T$_{3}$ & $\mathcal{U}(1000, 3000)$ & 2646$\pm$379 & K \\
       P$_{1}$ & $\mathcal{U}(-2, 3)$ & $-1.16 \pm 0.35$ & log bar \\
       P$_{2}$ & $\mathcal{U}(-6, -2)$ & $-2.46\pm0.24$ & log bar \\
       P$_{3}$ & -6 & ... & log bar \\
       $\rm [Fe/H]$ & $\mathcal{U}(-1, 3)$ & $1.51 \pm 0.44$ & dex \\
       C/O & $\mathcal{U}(0.1, 1.6)$ & $0.783 \pm 0.032$ & dex \\
    \hline
\end{tabular}
\end{table}

\begin{figure}[h!]
\centering
\includegraphics[width=6cm]{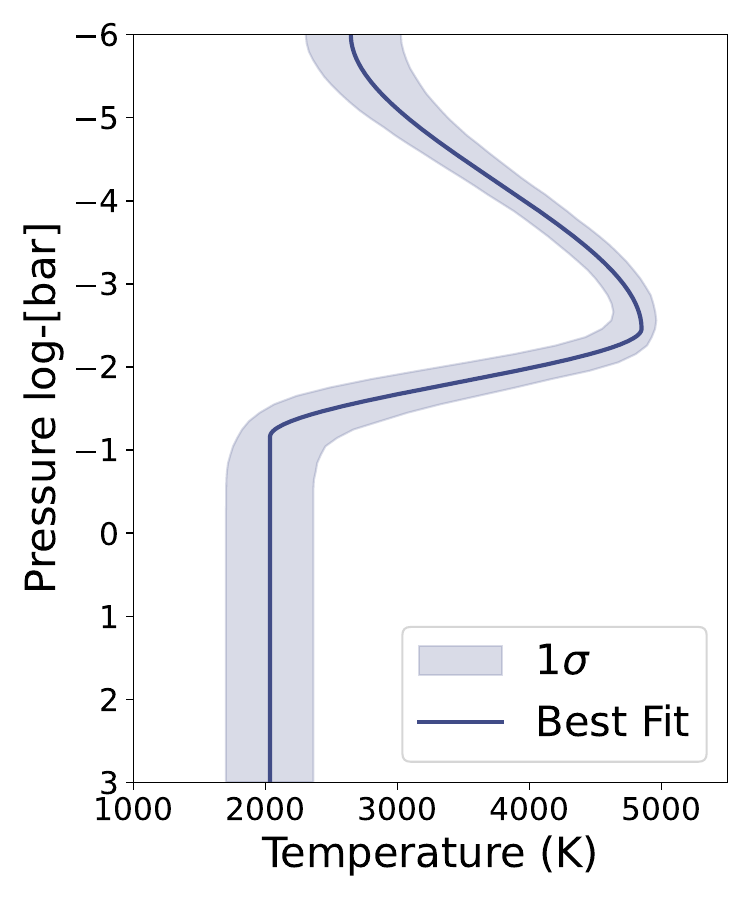}
\caption{Same as Fig. 13 but for the retrieved T-P profile from the EQ model.}
\label{fig:TP_EQ}
\end{figure}

EQ retrieval analysis yields consistent results with those from the HR spectroscopic study~\citep{changeat2022}. The best-fit model spectrum, the corresponding T-P profile, and the pressure-dependent chemical abundances are presented in Fig.\ref{fig:bestfit_eq}, Fig.\ref{fig:TP_EQ} and Table~\ref{table:eq_model_para}, respectively. The retrieved T-P profile suggests that the temperature inversion layer initiates near 1\,bar, consistent with previous results~\citep[e.g.][]{Nugroho2021, van2023, finnerty2023}, although the layer above the inversion is relatively cooler in our results. As shown in 
Table~\ref{table:eq_model_para}, the dayside atmosphere of WASP-33\,b should be highly metal-rich and exhibit a super-solar C/O ratio. 

\begin{figure}[h!]
        \centering
        \includegraphics[width=8cm]{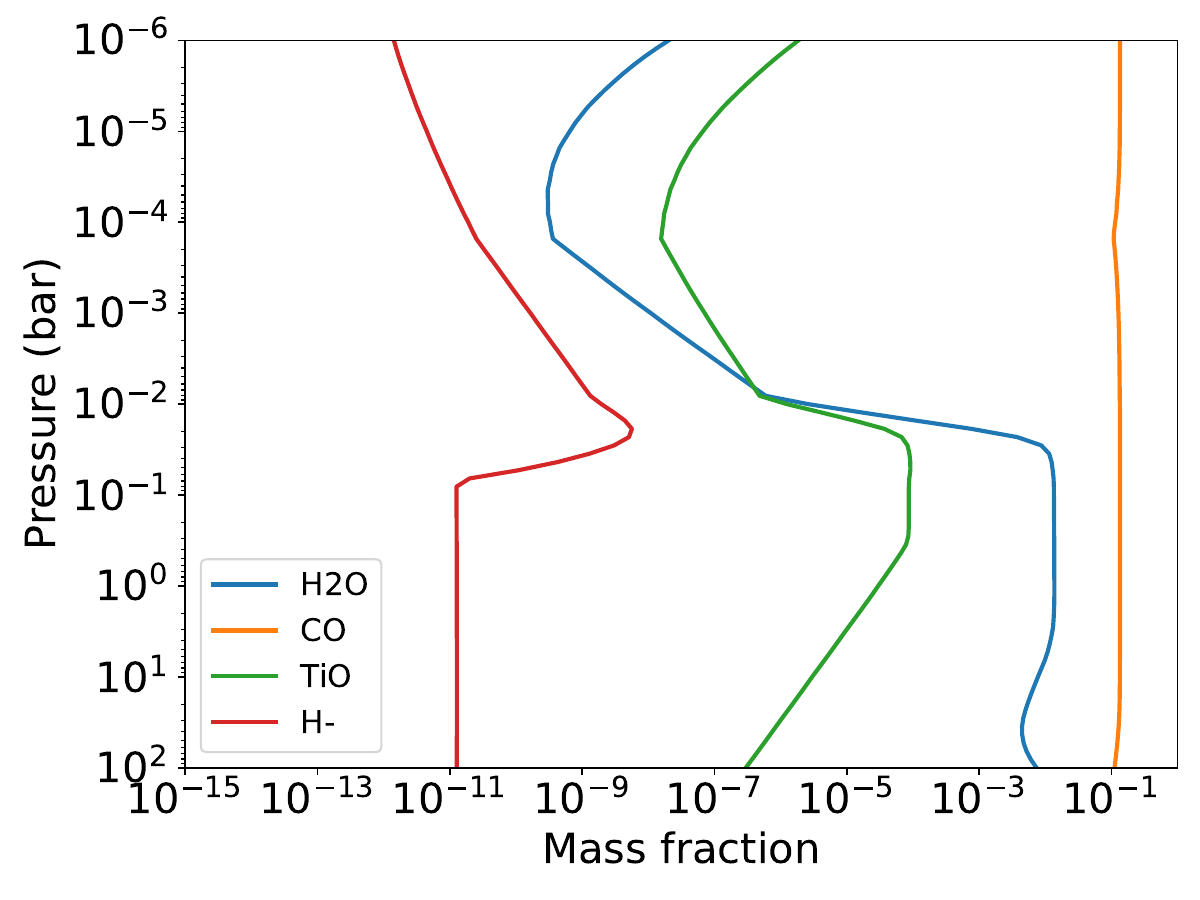}
        \caption{Mass mixing ratios of chemical species under EQ model as functions of atmospheric height. Apart from CO maintaining high concentrations and H$^{-}$ exhibiting a marked increase with rising temperature, all other molecules undergo thermal dissociation within the thermal inversion layer.}
        \label{fig:EQ_abundances}
   \end{figure}

Table~\ref{table:BE_EQ} summarizes the Bayesian-evidence validation for each species, showing strong support for H$^{-}$, H${2}$O, and CO, weak support for TiO, and no support for CH$_{4}$. Figure~\ref{fig:EQ_abundances} displays the MMRs of these species as a function of the pressure derived from the retrieved T-P profile. Given that CH$_{4}$ is disfavored by Bayesian evidence and is thermochemically implausible in the dayside atmosphere of UHJs, we therefore do not further examine the CH$_{4}$ MMR. Although H$^{-}$ abundance in the EQ model is appreciably lower than in the FREE model, the strong Bayesian evidence for H$^{-}$ in both frameworks underscores its necessity in the dayside atmosphere. The very high and nearly constant MMR of CO with altitude, together with the comparatively lower abundance of H$_{2}$O, is consistent with the Keck Planet Imager and Characterizer (KPIC) detection reported by \cite{finnerty2023}. The low concentration of TiO is likely a natural consequence of thermal dissociation in the hot upper atmosphere. Note that the retrieved emission spectrum is dominated by atmospheric layers between 1\,$\upmu$bar and 1\,bar; therefore, the species abundances within this pressure range are expected to be the most reliable.

\begin{table}
    \caption{Bayesian model comparison of the chemical species retrieved with EQ model.}  
    \renewcommand\arraystretch{1.4}
    \label{table:BE_EQ}    
    \centering                        
\begin{tabular}{l l l }      
    \hline
    \hline
       Species & $\Delta ln(\mathcal{Z})$ & Inferences \\
    \hline                      
       $\mathbf{H_{2}O}$ & \textbf{25.7} & \textbf{Strong} \\  
       $\mathbf{CO}$ & \textbf{36} & \textbf{Strong} \\
       CH$_{4}$ & -0.1 & Inconclusive \\
       $\mathbf{H^{-}}$ & \textbf{147.6} & \textbf{Strong} \\
       TiO & 2.5 & Weak \\
    \hline
\end{tabular}
    \tablefoot{Same as Table \ref{table:BE_free}, with uncertainties of $\sim$0.2.}
\end{table}

\subsection{EQ model retrieval}

In conventional planet formation theories, such as pebble accretion \citep{ormel2010, lambrechts2012}, metallicity and the C/O ratio generally show an inverse correlation, making it unlikely for planets to simultaneously possess high C/O ratios and high metallicity. However, the pebble drift theory proposed by \citet{booth2017} indicates that the metallicity of planets can be enriched not only by accreting solids but also by accreting metal-rich gas generated through the pebble drift mechanism. Synthesizing the dynamical and chemical properties of WASP-33\,b, our conclusions align with those outlined in \citet{finnerty2023}: the planet likely formed in a carbon- and solid-rich zone with high C/O ratio, such as near the CO$_{2}$ snow line ($\sim$10 au for an A-type primary star; \citealt{oberg2011,molliere2022}), accreted a high-metallicity atmosphere from its vicinity and migrated to its present location via a process like eccentric Lidov-Kozai effect \citep{naoz2011}. It should be noted, as suggested by \citet{finnerty2023}, that the observed high C/O ratio in the dayside atmosphere could also be biased due to data interpretation, namely by weak constraints on oxygen-bearing species. For example, our retrieval lacks constraints on OH, a photodissociation product of H$_{2}$O. Given that a significant amount of H$_{2}$O should be photodissociated into OH in upper atmospheres, this lack may lead to an overestimated C/O ratio.

On the other hand, using only metallicity and the C/O ratio as diagnostics is insufficient to accurately or unbiasedly constrain planet formation and evolution history \citep{feinstein2025}. Incorporating measurements of additional tracers (e.g., volatile elements and refractory element abundances) would better enrich the understanding of WASP-33\,b’s formation history. We recommend that future JWST observations of WASP-33\,b achieve more precise atmospheric constraints.

\section{Summary and conclusion}
\label{sec:conclusion}

We conducted two secondary-eclipse observations of the ultra-hot Jupiter WASP-33 b with CFHT/WIRCam to constrain the planet’s dayside atmosphere. To handle the thousands of iterations required for systematic correction and atmospheric retrievals, we developed a parallelized data reduction and analysis pipeline that substantially reduced computational time. After carefully removing the host star’s pulsation noise, we measured eclipse depths of $1565.2^{+228.6}_{-237.5}$\,ppm in the CO filter and $914.3^{+56.1}_{-57.0}$\,ppm in the CH4$_{\rm on}$ filter.

In addition to our CFHT/WIRCam observations, we incorporated HST/WFC3 spectroscopic data \citep{haynes2015} and Spitzer/IRAC photometric measurements -- 3.6\,$\upmu$m (\citealt{zhang2018}) and 4.5\,$\upmu$m \citep{dang2024} -- to jointly perform both free-chemistry and equilibrium-chemistry atmospheric retrievals using \texttt{petitRADTRANS}. Given the extreme dayside temperatures typical of ultra-hot Jupiters, we consider the equilibrium-chemistry model to be a more physically representative description of WASP-33\,b’s dayside atmosphere, compared to the free chemistry model. Accordingly, all results presented in the main findings in the following are based on the equilibrium model. Our main findings are summarized below.

\textbf{1. Thermal inversion and opacity sources.}
WASP-33\,b exhibits a clear thermal inversion from both free and equilibrium retrievals. Under the free-chemistry model, the dayside emission spectrum appears to be largely featureless, with Bayesian evidence strongly favoring the presence of H$^{-}$ opacity. However, the retrieved abundances of H$^{-}$ and e$^{-}$ are considered unreliable due to model limitations and spectral coverage. Incorporating blue or optical spectra in future joint retrievals will better constrain these species.

\textbf{2. High C/O ratio and metallicity in equilibrium chemistry.}
The equilibrium-chemistry retrieval yields a high C/O ratio of $0.78^{+0.03}_{-0.04}$ and a metallicity of $\sim 26 \times$ stellar, both more precisely constrained than in previous studies. The emission spectrum is primarily shaped by H$_2$O, CO, and H$^{-}$, while the existence of TiO remains uncertain due to weak Bayesian evidence. In the upper atmosphere, most molecules undergo thermal dissociation except CO.

\textbf{3. Implications for formation and migration.}
The combination of high metallicity and high C/O ratio is consistent with planetary enrichment through the accretion of metal-rich gas, as proposed by the pebble drift theory~\citep{booth2017}. A possible formation and migration scenario suggests that WASP-33\,b formed and/or accreted material in a carbon- and solid-rich region — possibly near the CO$_2$ snow line — before migrating inward via a mechanism such as the eccentric Lidov–Kozai effect \citep{naoz2011}. To test this hypothesis and rule out apparent carbon enrichment caused by weak constraints on oxygen-bearing species, we recommend follow-up JWST observations to obtain tighter constraints on additional elemental abundances.
WASP-33\,b remains an unusual exoplanet, orbiting a $\delta$\,Scuti variable star. Its high metallicity, elevated C/O ratio, and misaligned orbit all indicate a distinct formation and evolutionary history. Although stellar pulsations previously posed major challenges for measuring transit and eclipse depths, improved methodologies now allow for better mitigation of these signals and stronger isolation of the planetary component. Future observations—either at higher spectral resolution in targeted bands or with broader wavelength coverage at lower resolution—will enable more detailed characterization of WASP-33\,b’s atmosphere and provide a deeper insight into its formation and evolution.

\begin{acknowledgements}
We thank the referee and editor for their concise review and instructive suggestion. This research is supported by the National Key R\&D Program of China (2025YFE0102100, 2024YFA1611802, 2025YFE0213204), the National Natural Science Foundation of China under grant 62127901, 12588202, National Astronomical Observatories Chinese Academy of Sciences No. E4TQ2101, and the Pre-research project on Civil Aerospace Technologies No. D010301 funded by the China National Space Administration (CNSA), by the China Manned Space Program with grant no. CMS-CSST-2025-A16. We acknowledge financial support from the Agencia Estatal de Investigaci\'on of the Ministerio de Ciencia e Innovaci\'on MCIN/AEI/10.13039/501100011033 and the ERDF 'A way of making Europe' through projects PID2021-125627OB-C32 and PID2024-158486OB-C32. Q.Y.Zou is supported by the Program of China Scholarship Council (Grant CSC202510740003). MZ was supported by the Chinese Academy of Sciences (CAS), through a grant to the CAS South America Center for Astronomy (CASSACA) in Santiago, Chile. This research uses data obtained through the Telescope Acess Program(TAP), which has been funded by the National Astronomical Observatories of China, the Chinese Academy of Sciences, and the Special Fund for Astronomy from the Ministry of Finance. Two nights (Oct 25\& Dec 24,2015) with WIRCam on the 3.6m CFHT telescope were distributed to us for scientific studies of exoplanetary atmosphere via the TAP. We also thank Dr. Jie Zheng, Jingyuan Zhao, Hengkai Ding and Yingjie Cai for their constructive comments on this work.
\end{acknowledgements}

\bibliographystyle{aa}
\bibliography{references.bib}

\end{document}